\documentclass[12pt,onecolumn,floatfix,superscriptaddress,tightenlines,nofootinbib]{revtex4-1}
\pdfoutput=1
\usepackage[colorlinks=true,citecolor=blue,linkcolor=blue,breaklinks=true]{hyperref}
\usepackage{amsmath,amssymb}
\usepackage{epsfig} 
\usepackage{graphicx}        
\usepackage{url}
\usepackage{color}
\usepackage{multirow}
\usepackage{placeins}
\usepackage[dvipsnames]{xcolor}
\usepackage{braket}
\hypersetup{
  colorlinks=true,
  linkcolor=red,
  filecolor=magenta,   
  urlcolor=blue,
  citecolor=blue,
}

\allowdisplaybreaks

\bibpunct{[}{]}{,}{n}{}{,}

\def\beq{\begin{equation}}
\def\eeq{\end{equation}}
\def\bea{\begin{eqnarray}}
\def\eea{\end{eqnarray}}
\makeatletter
\@addtoreset{equation}{section}

\begin{document}

\title{Baryogenesis from  ultralight primordial black holes and strong gravitational waves from cosmic strings}
\author{Satyabrata Datta}
\email{satyabrata.datta@saha.ac.in}
\affiliation{Saha Institute of Nuclear Physics, HBNI, 1/AF Bidhannagar,
Kolkata 700064, India}
\author{Ambar Ghosal}
\email{ambar.ghosal@gmail.com}
\affiliation{Saha Institute of Nuclear Physics, HBNI, 1/AF Bidhannagar,
Kolkata 700064, India}
\author{Rome Samanta}
\email{romesamanta@gmail.com}
\affiliation{CEICO, Institute of Physics of the Czech Academy of Sciences, Na Slovance 1999/2, 182 21 Prague 8, Czech Republic}


\begin{abstract} 
Ultralight primordial black holes (PBHs)($\lesssim10^9$g) completely evaporate via Hawking radiation (HR) and produce all the particles in a given theory regardless of their other interactions. If the right handed (RH) neutrinos are produced from PBH evaporation, successful baryogenesis via leptogenesis predicts mass scale of  RH neutrinos as well as black holes. We show that, given the lepton number violation (generation of RH neutrino masses) in the theory is a consequence of a gauged $U(1)$ breaking which is then followed by the formation of PBHs,  a network of cosmic strings naturally gives rise to  strong stochastic gravitational wave (GW) signal at the sensitivity level of pulsar timing arrays (PTA) and LIGO5.  Besides,  due to a  transient period of black hole domination in the early universe, for which baryon asymmetry is independent of initial PBH density, a  break in the GW spectra occurs around MHz frequency. Therefore, to observe the break along with the usual GW signal by the emission of gravitons via HR,   GW detectors at higher frequencies are called for. The recent finding by the NANOGrav PTA of a stochastic common spectrum process (interpreted as GWs) across many pulsars is in tension with PBH baryogenesis  for large cosmic string loops ($\alpha\simeq 0.1$).

\end{abstract}

\maketitle

\section{Introduction}
Physics of Primordial Black Holes (PBH)\cite{bh1,bh2,bh3} that entails rich cosmological consequences, has gained a renewed interest after the discovery of Gravitational Waves (GW) from black hole mergers by LIGO and Virgo collaboration \cite{gw1,gw2,gw3,gw4,gw5,gw6,gw7}. Starting from the mass $M_{BH}\simeq 0.1 $g up to few hundreds of solar mass, PBHs constitute seemingly  intriguing physical aspects. Long-lived PBHs with mass ($M_{BH}\gtrsim 10^{15} $g) that survive until today may contribute to a significant portion of Dark Matter (DM)\cite{pdm1,pdm2,pdm3,pdm4,pdm5,pdm6,pdm7,pdm8,pdm9,Miller:2020kmv,Cai:2018dig}, whereas PBHs with $M_{BH}\lesssim 10^9$g  completely evaporate before Big Bang Nucleosynthesis (BBN) ($T\sim4 $ MeV) via Hawking Radiation (HR)\cite{hr}. On the other hand, the intermediate mass black holes $10^9{\rm g}\lesssim M_{BH}\lesssim 10^{15}$g are subjected to several constraints, e.g., from BBN \cite{bbn1,bbn2,bbn3}. Though the ultralight PBHs ($0.1 {\rm g}\lesssim M_{BH}\lesssim 10^9$g) completely evaporate,  the particle emission being democratic in nature and poor constraints on initial PBH density at the formation\footnote{can be constrained, e.g., with  the production of gravitinos in supersymmetric models\cite{khlo1,khlo2}.}, they leave plenty of room that can accommodate new physics. Perhaps this is why parallel to the usual study of PBH as Dark matter (DM)  ($\gtrsim 10^{15}$g), physics of ultralight PBHs have gained a lot of interest in recent time. This includes, for example, production of DM particles\cite{dm0,dm1,dm2,dm3,dm4,dm4a,dm5,dm5a,dm6,dm7,dm8,dm9,dm10,dm11,Kitabayashi:2021hox}, generation of baryon asymmetry\cite{br0,br0a,dm0,br1,br2,br3,br4,Aliferis:2020dxr}, change in GW spectrum induced by scalar perturbations\cite{yana}, GWs induced by ultralight
 primordial black holes\cite{Papanikolaou:2020qtd,bbn4} and study of vacuum stability of the Standard Model (SM) Higgs\cite{hd1,hd2}. In this work, we confine ourselves with the detailed study of baryogenesis via non-thermal leptogenesis by ultralight PBH evaporation and its possible test with GWs.  \\

Amongst several mechanisms of baryogenesis, the simplest one is baryogenesis via leptogenesis\cite{lep0,lep0a,lep0b,lep1,lep2,lep3,lep4,lep5,lep6} from the decays of heavy Right Handed (RH) neutrinos which are introduced in the SM to obtain light neutrino masses via Type-I seesaw mechanism\cite{see1,see2,see3,see4,see5}. However, to test standard leptogenesis within Type-I seesaw, one has to either lower the RH mass scales for collider searches\cite{Ars,Ham, dev1,dev2} or invoke restrictions on the relevant parameters, e.g., considering discrete symmetries\cite{fasy1,fasy2,fasy3,fasy4} and theories like SO(10) grand unification (GUT)\cite{so1,so2,so3,so4,so5}. Recently, a pathway to probe leptogenesis with GWs has been introduced in Ref.\cite{lepcs1}. The idea is based on formation of cosmic  string network\cite{cs1,cs2,cs3} upon breaking of an Abelian symmetry and consequent emission of GWs. Though emission of GWs from cosmic string loops still remains under debate,  numerical simulations  based on the Nambu–Goto action\cite{ng1,ng2} indicate  that cosmic string loops loose energy dominantly via GW radiation, if the underlying broken symmetry corresponds to a local gauge symmetry, e.g., $U(1)$, which in the context of seesaw, can be readily promoted to a $U(1)_{B-L}$\cite{Davidson:1978pm,moha1,moha2}. Once $U(1)_{B-L}$ gets spontaneously broken, RH neutrinos get mass and cosmic strings are generated. The scale of leptogenesis (RH masses) and amplitude of the GWs, both are related to the symmetry breaking scale ($\Lambda_{CS}$). Therefore, one may probe the scale of leptogenesis through the detection of GWs.  With this idea, two other variations of leptogenesis have also been studied\cite{lepcs2,lepcs3}. In fact, due to  the most distinguishable feature of strong GW signal from cosmic strings across a wide range of frequencies, there is a rapidly growing interest in this field\cite{gr1,gr2,gr3,gr4,gr5,gr6,gr7,gr8,gr9,gr10,gr11,gcs1,gcs2,Fornal:2020esl}. Moreover, the recent finding of stochastic common spectrum processes across many pulsars by the NANOGrav collaboration\cite{nanog} has strengthened the possibility to consider GW emission from cosmic strings, since the new data can be explained better with cosmic string models\cite{lepcs3,nan1,nan2,nan3,nan4,nan5,nan6} than the single value power spectral density as suggested by the models of supermassive black holes (SMBHs). 
 
In this work, we follow a similar idea as in Ref.\cite{lepcs1,lepcs2,lepcs3} to study `GW friendly leptogenesis (dynamical generation of RH mass scale)' from PBH evaporation wherein massive RH neutrinos are emitted by the PBHs and then decay CP asymmetrically to create lepton asymmetry\cite{lep1}. However, as we explain below, unlike Ref.\cite{lepcs1}, here we give more emphasis on the PBH mass rather than RH neutrino masses. We discuss a scenario where after the end of inflation, there is a $U(1)_{B-L}$ phase-transition (at $\Lambda_{CS}$) so that RH neutrinos become massive\cite{lepcs1}  which is then followed by the formation of PBHs (at $T_{Bf}$) that  emit massive RH neutrinos thereafter, i.e., $\Lambda_{CS}\gtrsim T_{Bf}$. In this scenario, i.e., in the non-thermal leptogenesis from PBH evaporation,  the requirement to obtain correct baryon asymmetry, strongly constrains RH neutrino masses as well as the mass of the PBHs. The most interesting outcome is,  one obtains an upper bound on $M_{BH}$ which translates into a lower bound on the formation temperature ($T_{Bf}\gtrsim 10^{15}$ GeV) of PBHs and hence on $\Lambda_{CS}$. Since with the decrease of $M_{BH}$, the formation temperature increases\cite{pdm6}, one naturally obtains GWs with large amplitude throughout the parameter space of a successful leptogenesis, given the fact that the GW amplitude increases with the breaking scale $\Lambda_{CS}$\cite{gr2} (for example, $\Lambda_{CS}\gtrsim 10^{15}$ GeV would correspond to $\Omega_{GW}h^2(f)\gtrsim 10^{-9}$). Therefore, the mechanism is fully testable in most of the planned and present GW detectors over a wide range of frequencies. We focus more on the case in which after the formation, PBHs dominate the energy density of the universe until they completely evaporate, which is also motivated by the fact that in this case, the baryon asymmetry is independent of PBH density at formation\cite{br0}. With such a non-standard cosmological evolution (a matter domination before the usual radiation domination) one would expect a  break in the GW spectrum\cite{gr2,lepcs2}. We show that, for a successful leptogenesis scenario, this break happens at much higher frequencies compare to the sensitivity,  e.g, of LIGO ($\sim 25$ Hz). Finally, we discuss the relevance of PBH baryogenesis in the light of recent 12.5 yrs NANOGrav PTA data and conclude that the predicted GW amplitude for large cosmic string loops $\alpha^{max}\simeq 0.1$\cite{csa1,csa2} is in tension with the NANOGrav data.\\

The rest of the paper is organised as follows: In Sec.\ref{sec2}, we discuss the dynamics of PBH energy density  and its impact on thermal leptogenesis. Sec.\ref{sec3} is devoted to the discussion on baryogenesis via non-thermal leptogenesis in radiation domination and  in PBH domination. In Sec.\ref{sec4}, we review the GW emission from cosmic string and present the numerical results which also include the fitting to the  12.5 yrs NANOGrav data. We conclude in Sec.\ref{sec5}.
\section{PBH dynamics: Radiation domination Vs. Black hole domination}\label{sec2}
In this section, we first derive the condition on PBH mass and its density at the formation  to understand the scenarios like PBH domination and radiation domination after the time of black hole formation (Ref.\cite{pdm6}: a review that discusses PBH formation mechanisms).  For the radiation energy density to dominate the universe until black hole evaporation,  the ratio of the energy density of black holes and the radiation at the time of black hole evaporation  would satisfy 
\bea
r(t_{ev})=\frac{\rho_{ BH}(t_{ev})}{\rho_{\rm R}(t_{ev})}<1,\label{rad dom}
\eea
where $t_{ev}$ is the time at evaporation, $\rho_{ BH}$ and $\rho_{R}$ are the black hole and radiation densities respectively.
Thus the ratio of the $r$ parameters at the time of  black hole formation and evaporation can be calculated as
\bea
\frac{r(t_{ev})}{r(t_{Bf})}=\frac{a(t_{ev})}{a(t_{Bf})}=\left(\frac{t_{ev}}{t_{Bf}}\right)^{1/2}\label{rr}
\eea
with $t_{Bf}$ being the time at black hole formation.
Recalling the Friedmann equation 
\bea
H(t)^2=\frac{8\pi}{3M_{Pl}^2}\rho_{R}(t),~~ {\rm with}~~M_{Pl}=1.22\times 10^{19}~{\rm GeV}\label{frd1}
\eea
and using the Hubble parameter in radiation domination and the radiation density as
\bea
H=\frac{1}{2t}, ~~\rho_{R}(T)=\frac{\pi^2 g_*(T) T^4}{30},\label{radeng}
\eea

where $g_*(T)$ ($\simeq$ 106.75 in SM) is the effective degrees of freedom that contribute to the radiation,  Eq.\ref{rr} can be re-written as 
\bea
\frac{r(t_{ev})}{r(t_{Bf})}=\left(\frac{g_*(T_{Bf})}{g_*(T_{ev})}\right)^{1/4}\frac{T_{Bf}}{T_{ev}}.\label{key}
\eea
Therefore Eq.\ref{rad dom} translates to the condition
\bea
r(t_{Bf})\equiv \beta <\left(\frac{g_*(T_{ev})}{g_*(T_{Bf})}\right)^{1/4}\frac{T_{ev}}{T_{Bf}}.\label{condrad}
\eea
We now need to calculate the black hole formation temperature ($T_{Bf}$) and the evaporation temperature $(T_{ev})$. The formation temperature can be calculated once we know the formation time $t_{Bf}$.  Assuming radiation domination at black hole formation, the mass of the black hole from gravitational collapse is typically close to the value enclosed by the post-inflation particle horizon and is given by\cite{pdm6}
\bea
M_{BH}=\gamma\frac{4}{3}\pi(H_{Bf}^{-1})^3 \rho_{Bf}~~{\rm with}~~\rho_{Bf}=\frac{3 H_{Bf}^2 M_{Pl}^2}{8\pi}, ~~H_{Bf}=\frac{1}{2t_{Bf}}.
\eea
The quantity $\gamma\simeq 0.2$ is a numerical factor which depends on the details of the gravitational collapse mechanism. 
Thus the time $t_{Bf}$ can be calculated as
\bea
t_{Bf}=\frac{M_{BH}}{M_{Pl}^2\gamma}.
\eea
Now using the Friedmann equation (Eq.\ref{frd1}) and radiation energy density (Eq.\ref{radeng}) one obtains the black hole formation temperature as
\bea
T_{Bf}=\left(\frac{45 \gamma^2}{16\pi^3 g_*(T_{Bf})}\right)^{1/4}\left(\frac{M_{Pl}}{M_{BH}}\right)^{1/2}M_{Pl}.\label{tbf}
\eea
We now calculate the black hole evaporation temperature $T_{ev}$.  Assuming radiation dominated universe throughout the  evolution of the black holes, we can write the Hubble parameter as 
\bea
H(t_{ev })^2=\frac{1}{4t_{ev}^2}\simeq \frac{1}{4\tau^2},
\eea
where the $\tau$ is the lifetime of the PBH. Therefore, from the Friedmann equation in Eq.\ref{frd1} and the expression for the radiation energy density in Eq.\ref{radeng}, we can calculate the temperature $T_{ev}$ as 
\bea
T_{ev}=\left(\frac{45 M_{Pl}^2}{16\pi^3 g_*(T_{ev})\tau^2}\right)^{1/4}.\label{teva}
\eea
The lifetime of the PBH can be calculated from the dynamics of the mass loss of a PBH via Hawking radiation\cite{hr}. The rate at which PBH looses its mass is given by 
\bea
-\frac{dM_{BH}}{dt}=f_{ev}(4\pi r_{BH}^2)\frac{dE}{dt},\label{hwkrad}
\eea
where the rate of energy loss can be calculated as
\bea
\frac{dE}{dt}=2g_{BH}(T_{BH})\pi^2\int_0^\infty d\nu\frac{\nu^3}{{\rm exp} (2\pi\nu/T)-1}=\frac{\pi^2}{120}g_{BH}(T_{BH}) T_{BH}^4.
\eea
The quantity $f_{ev}$ is the efficiency of PBH evaporation and $r_{BH}$ is the Schwarzschild radius of the PBH given by 
\bea
r_{BH}=2GM_{BH}.\label{sc}
\eea
The quantity $g_{BH}(T_{BH})$ counts the bosonic and fermionic degrees of freedom below $T_{BH}$. 
\begin{figure}
\includegraphics[scale=.35]{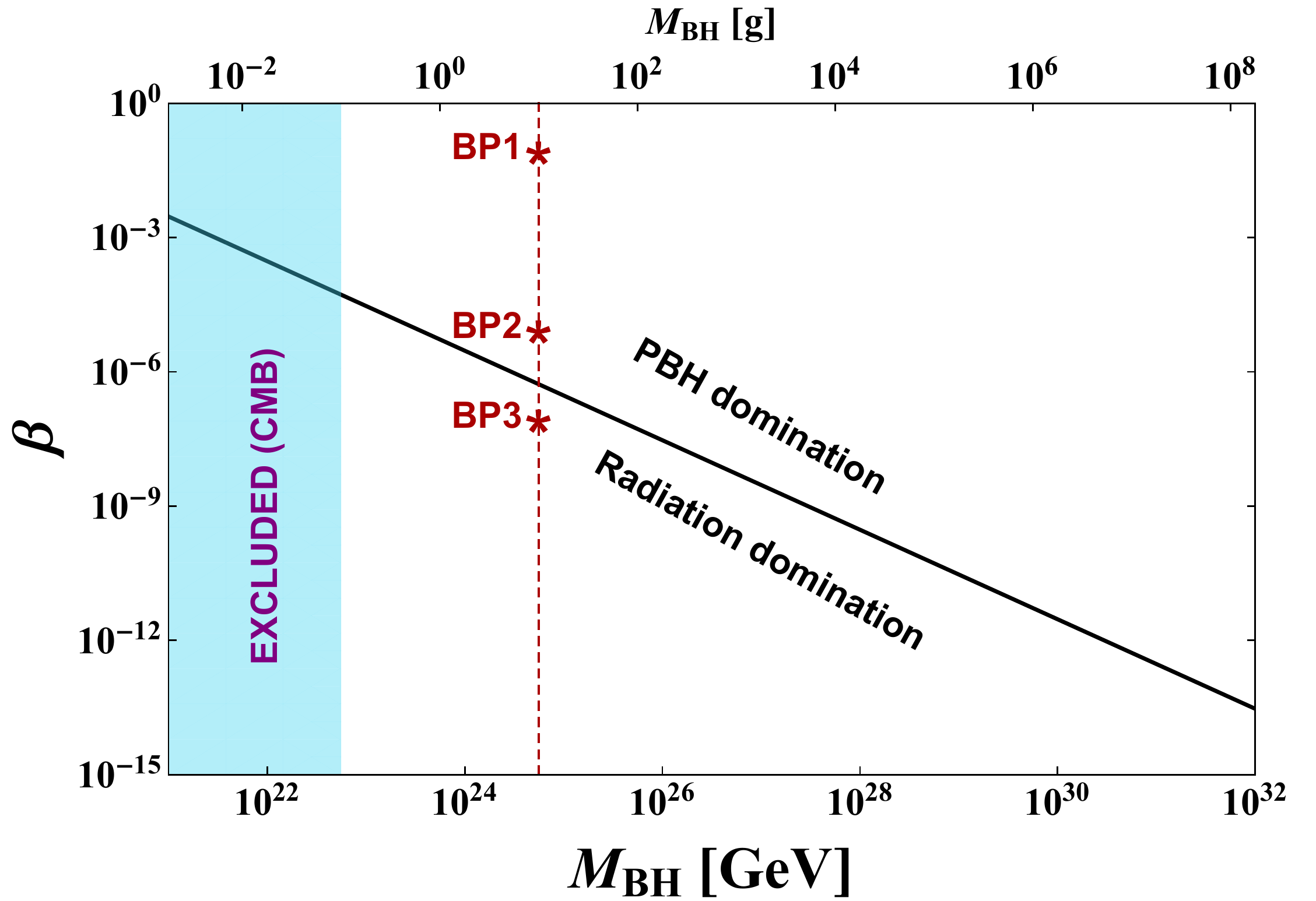}\includegraphics[scale=.35]{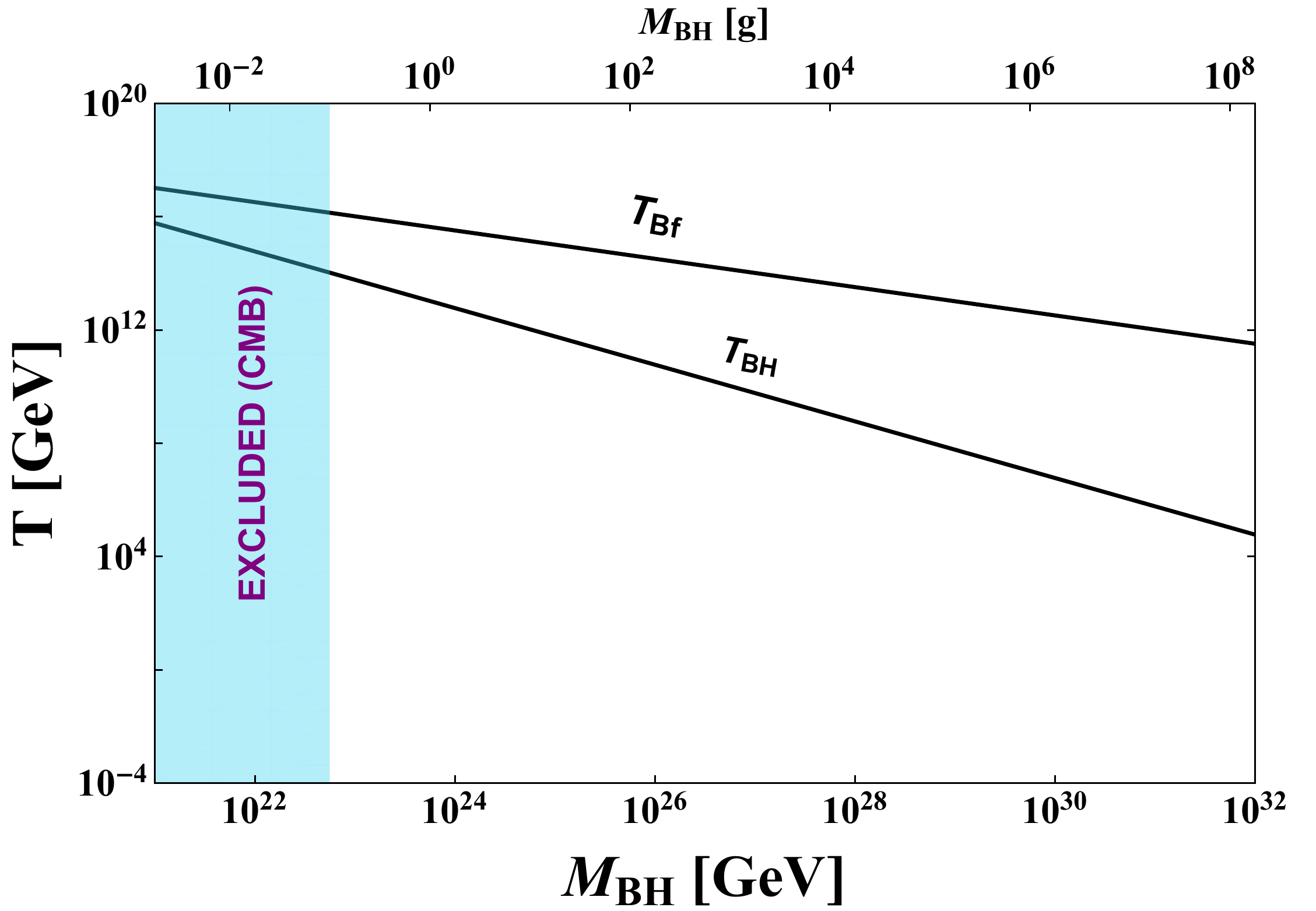}
\caption{Left: $\beta$ vs. $M_{BH}$. The  solid black line separates the regions of PBH domination and radiation domination. With the red stars, we indicate three benchmark points for which we show the evolution of the energy densities by solving the Boltzmann equations.  Right: PBH formation temperature and the Hawking temperature as a function of PBH mass. In each of the plots, the light blue region is excluded by the constraint from CMB on Hubble scale of inflation $H_{\rm inf}\lesssim 3\times 10^{14}$ GeV at 95$\%$ CL\cite{Planck}.}\label{fig1}
\end{figure}
Recalling the PBH temperature ($T_{BH}=M_{Pl}^2/8\pi M_{BH}$\cite{hr}) and using the  expression for $r_{BH}$, Eq.\ref{hwkrad} can be recast as 
\bea
\frac{dM_{BH}}{dt}=-\frac{\mathcal{G} g_{*B}(T_{BH})}{30720\pi}\frac{M_{Pl}^4}{M^2_{BH}},
\eea
where we have used $f_{ev}g_{BH}(T_{BH})=\mathcal{G}g_{*B}(T_{BH})$ with $\mathcal{G}\simeq 3.8$ being the graybody factor and $g_{*B} \simeq 100$ (counts all the particle species below $T_{BH}$) in the SM with three left handed neutrinos\cite{MacGibbon:1991tj}. It is now trivial to find the lifetime $\tau$ as
\bea
\tau=\int_{t_{Bf}}^{t_{ev}}dt=-\int_{M_{BH}}^0d M_{BH}\frac{30720\pi M_{BH}^2}{\mathcal{G}g_{*B}(T_{BH})M_{Pl}^4}=\frac{10240\pi M_{BH}^3}{\mathcal{G}g_{*B}(T_{BH})M_{Pl}^4}.\label{lt}
\eea
Therefore, combining Eq.\ref{lt}, Eq.\ref{teva} and Eq.\ref{tbf}, we can recast Eq.\ref{condrad} as 
\bea
\beta<\gamma^{-1/2}\left(\frac{\mathcal{G }g_{*B}(T_{BH})}{10240\pi}\right)^{1/2}\frac{M_{Pl}}{M_{BH}}.\label{beta}
\eea
On the left-hand side of figure \ref{fig1}, the relative PBH density $\beta$ as a function of PBH mass has been shown. As one sees, for very light PBHs to dominate the energy density, the relative PBH density at formation should be much higher. Thus in the context of gravitino production\cite{khlo1,khlo2} very light PBH domination may not be possible. On the right-hand side, we plot the PBH formation temperature and the PBH temperature as a function of PBH mass. From this figure one infers that due to the chosen breaking scale of $U(1)_{B-L}$, i.e., $\Lambda_{CS}\gtrsim T_{Bf}$, the RH mass scale could be higher as well as lower than $T_{BH}$, depending upon the perturbative Yukawa coupling ($y_{N_i}N_{iR}N_{iR}\phi_{B-L}$) of the $U(1)_{B-L}$ scalar and the RH neutrino fields. Therefore, in the computation of leptogenesis, we need to consider RH neutrino production from PBH in both the scenarios, i.e.,  $T_{BH}>M_R$ as well as $T_{BH}<M_R$.
To understand the dynamics of black hole vs. radiation energy density throughout the evolution of the universe, one needs to solve the following Boltzmann equations\cite{Giudice:2000ex}
\bea
\frac{d\rho_R}{dt}+4H\rho_R=-\frac{\dot{M}_{BH}}{M_{BH}}\rho_{BH},\label{be1}\\
\frac{d\rho_{BH}}{dt}+3H\rho_{BH}=+\frac{\dot{M}_{BH}}{M_{BH}}\rho_{BH},\label{be2}\\
\frac{ds}{dt}+3Hs=-\frac{\dot{M}_{BH}}{M_{BH}}\frac{\rho_{BH}}{T},\label{be3}
\eea
where Eq.\ref{be3} represents the non-conservation of entropy due to black hole evaporation and $\dot{M}_{BH}\equiv dM_{BH}/dt$. It is convenient to recast the above set of equations in terms of temperature. From Eq.\ref{be3}, the relation between temperature and time can be derived as
\bea
\frac{1}{T}\frac{dT}{dt}=-\left(H+\frac{1}{3g_{*s}(T)}\frac{dg_{*s}(T)}{dt}+\frac{\dot{M}_{BH}}{M_{BH}}\frac{\rho_{BH}}{4\rho_{R}}\right)=-\tilde{H}.\label{temvar}
\eea
where  we have used 
\bea
s(t)=\frac{2\pi^2 g_{*s} T^3}{45}\label{entr}
\eea
and in the right-hand side of Eq.\ref{temvar}, we have introduced the  quantity $\rho_R$ after considering $g_*(T)\equiv g_{*s}(T)$. Using Eq.\ref{temvar}, Eq.\ref{be1} and Eq.\ref{be2} can be  re-written as 
\bea
\frac{d\rho_{R}}{dz}+\frac{4}{z}\rho_R&=&0\label{den1}\\
\frac{d\rho_{BH}}{dz}+\frac{3}{z}\frac{H}{\tilde{H}}\rho_{BH}&-&\frac{\dot{M}_{BH}}{M_{BH}}\frac{1}{z\tilde{H}}\rho_{BH}=0,\label{den2}
\eea
where $z=M_0/T$ with $M_0$ being an arbitrary mass scale which we consider as $M_0\equiv T_{Bf}$. On the left-hand side of figure \ref{fig2}, we show the  normalised energy densities of radiation and PBHs for different values of $\beta$, as indicated in figure \ref{fig1}. We take a benchmark value $M_{BH}=10$ g for computation. It is clear that with the increase of $\beta$, one gets a larger epoch of PBH domination which has a great impact on the thermally produced lepton asymmetry. As we mentioned in the introduction, we shall consider only non-thermal production of RH neutrinos by PBH evaporation. This requires a justification that the non-thermal contribution from PBH dominates over the thermal contribution from the plasma. 

\begin{figure}
\includegraphics[scale=.35]{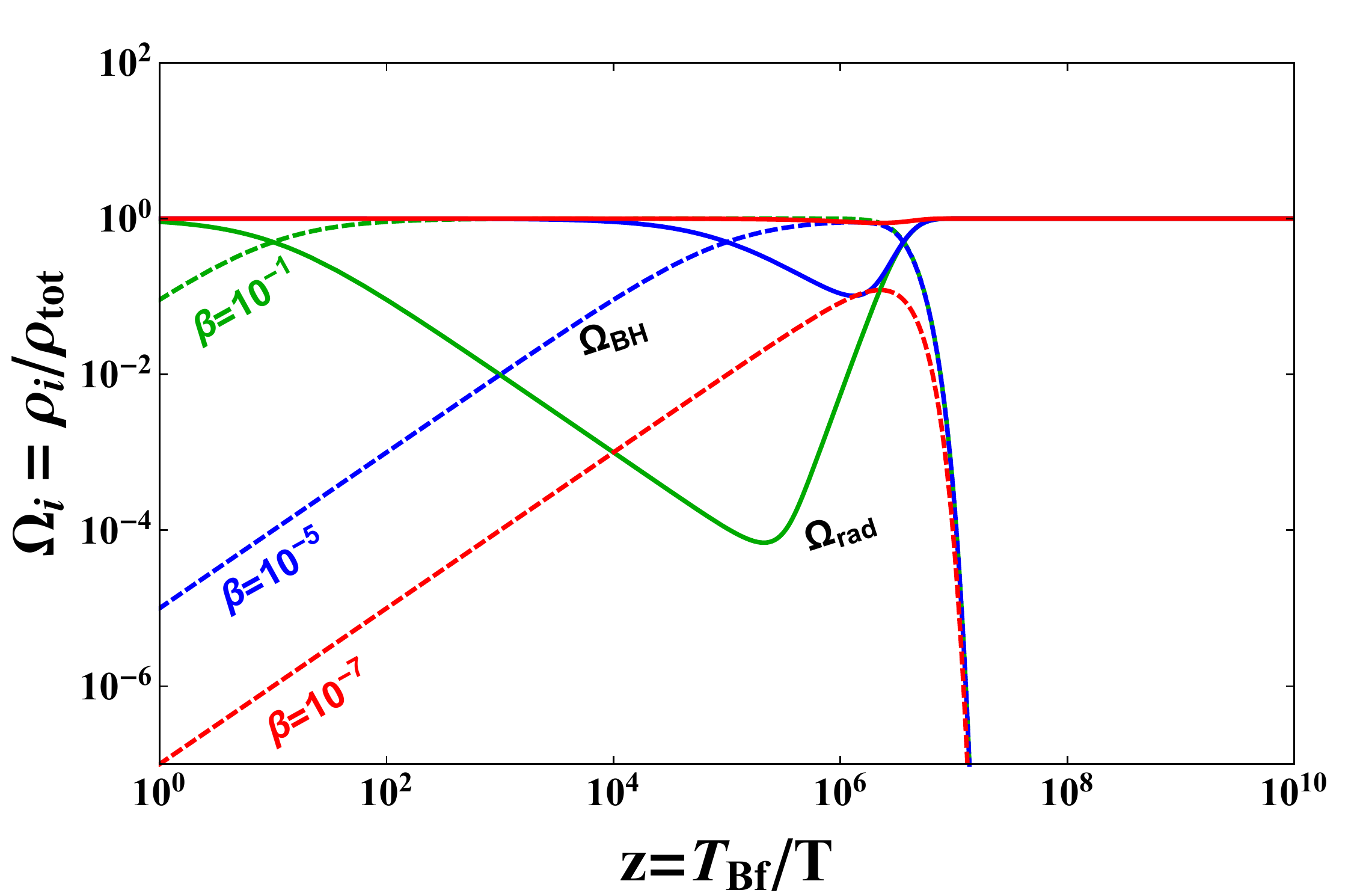} \includegraphics[scale=.5]{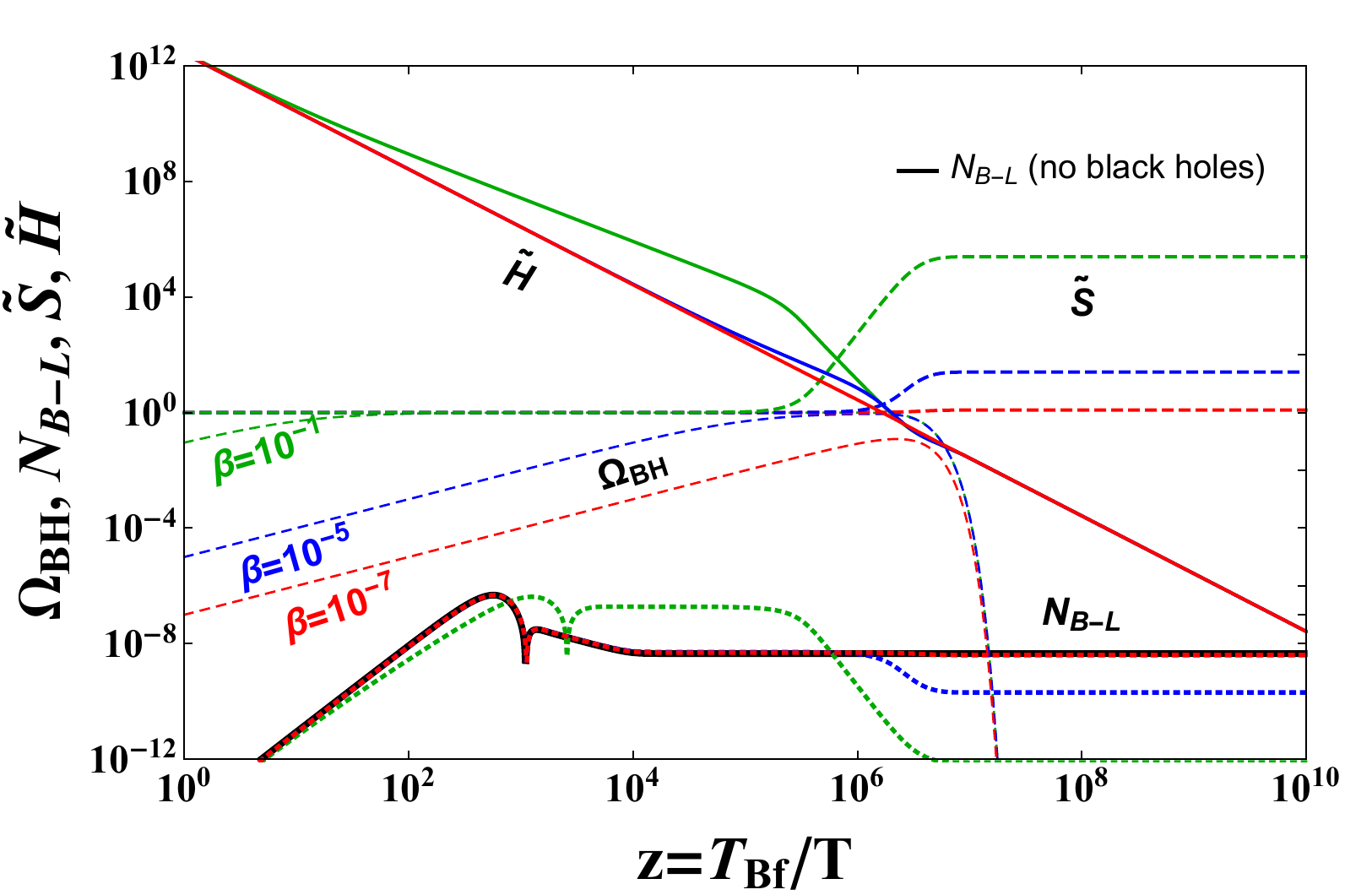}
\caption{Left: Evolution of radiation and black hole energy densities (normalised to the total energy density). Right: A representative figure to show the washout (via entropy production by PBH evaporation) of the thermally produced lepton asymmetry (solid line). The solid  black line representing $N_{B-L}$ indicates the thermally produced asymmetry with no PBHs. We choose $\Delta=10^{-3}$, $\varepsilon\sim10^{-6}$ and $K_1=50.$}\label{fig2}
\end{figure}
As we explain below, this is indeed the case for a significant period of PBH domination (large $\beta$). In presence of PBHs, using the time-temperature relation in Eq.\ref{temvar}, the standard Boltzmann equations (BEs) for thermal leptogenesis\cite{lep4} ($N_1$-dominated scenario and without flavour effects\cite{fle1,fle2,fle2a,fle3,fle4,fle5}) can be generalised to 
\bea
\frac{dN_{N_1}}{dz}&=&\frac{3}{\tilde{H} z}(\tilde{H}-H)N_{N_{1}}-\frac{\Gamma_{D_{N_1}}}{\tilde{H} z}(N_{N_1}-N_{N_1}^{\rm eq}),\label{bhlep1}\\
\frac{dN_{B-L}}{dz}&=&-\varepsilon_1\frac{\Gamma_{D_{N_1}}}{\tilde{H} z}(N_{N_1}-N_{N_1}^{\rm eq})-\left(\frac{\Gamma_{ID_{N_1}}}{\tilde{H} z}-\frac{3}{\tilde{H}z}(\tilde{H}-H)\right)N_{B-L},\label{bhlep2}
\eea
where 
\bea
\Gamma_{D_{N_1}}=K_1\left(\frac{m^*}{8\pi v^2}\right)\Delta^2 T_{Bf}^2\frac{\mathcal{K}_1(z_1)}{\mathcal{K}_2(z_1)},~~\Gamma_{ID_{N_1}}=\frac{1}{4}K_1\left(\frac{m^*}{8\pi v^2}\right)\Delta^2 T_{Bf}^2z_1^2\mathcal{K}_1(z_1)\label{neb}
\eea

In Eq.\ref{neb}, $N_{N_i}$s are number densities normalised to the ultra-relativistic number density of $N_1$ ($\propto T^3$),  $\mathcal{K}_i$s are the modified Bessel functions, $N_{N_1}^{\rm eq}=\frac{1}{2}z_1^2\mathcal{K}_2(z_1)$ and $\Delta=M_1/T_{Bf}$ which implies  $z_1(=M_1/T)=z\Delta$. The quantity $K_1$ depends on the Yukawa couplings of SM Higgs, lepton doublets and $N_1$ which in the strong washout scenario has a value ranging typically from $K_1\sim 3-50$\cite{lep4}. The equilibrium neutrino mass is given by $m^*\simeq 10^{-3}$ eV. The quantity $\varepsilon$ is the CP violation responsible for leptogenesis and has an upper bound in the hierarchical limit of the RH neutrinos as (see Ref.\cite{Samanta:2020gdw} for a derivation)
\bea
\varepsilon_1 \lesssim \frac{3M_1 m_3}{8\pi v^2},\label{upep}
\eea
where $v=174$ GeV is the vacuum expectation value of the SM Higgs  and we assume normal mass ordering of the light neutrinos with $m_3$ being the heaviest light neutrino. Note in Eq.\ref{bhlep1} and Eq.\ref{bhlep2}, that in the limit $\tilde{H}\rightarrow H$ we recover the standard BEs for thermal leptogenesis\cite{lep4} (no black holes), i.e.,  
\bea
\frac{dN_{N_1}}{dz}&=&-D_{N_1}(N_{N_1}-N_{N_1}^{\rm eq}),\label{bhlep1a}\\
\frac{dN_{B-L}}{dz}&=&-\varepsilon_1 D_{N_1}(N_{N_1}-N_{N_1}^{\rm eq})-W^{ID}_{N_1}N_{B-L},\label{bhlep2a}
\eea
where
\bea
\frac{\Gamma_{D_{N_1}}}{H z}\equiv D_{N_{N_1}}=K_1\Delta^2 z \frac{\mathcal{K}_1(z_1)}{\mathcal{K}_2(z_1)}, ~~~\frac{\Gamma_{ID_{N_1}}}{H z}\equiv W^{ ID}_{N_1}=\frac{1}{4}K_1\Delta^4 z^3 \mathcal{K}_1(z_1).\label{deide}
\eea
The main effect that is caused by introducing PBHs, that in Eq.\ref{bhlep2}, in addition to the standard inverse decays ($\Gamma_{ID_{N_1}}$), one has a new `washout term' which at the time of PBH evaporation could be strong enough  to erase any asymmetry produced by the thermal plasma. We solve the four coupled equations presented in Eq.\ref{den1}, Eq.\ref{den2}, Eq.\ref{bhlep1} and Eq.\ref{bhlep2} to understand the dynamics of the thermally produced lepton asymmetry in presence of PBHs. On the right hand side of figure \ref{fig2}, it can be seen that as one goes from a radiation dominated phase to a PBH dominated phase (i.e., when $\beta$ increases), the contribution from the thermal plasma becomes negligible. For $\beta=10^{-7}$, the red dotted line and the black solid line (solution of Eq.\ref{bhlep1a} and Eq.\ref{bhlep2a}) practically coincide. This means for lower values of $\beta$, the effect of PBHs is negligible, as expected.  The produced entropy can be calculated by solving
\bea
\frac{da}{dz}=\left(1-\frac{\dot{M}_{BH}}{M_{BH}}\frac{\rho_{BH}}{4\rho_{R}\tilde{H}}\right)\frac{a}{z},
\eea
where we have neglected the time variation of $g_{*s}$.
On the right-hand side of figure \ref{fig2}, the quantity  $\tilde{S}\propto a^3/z^3$ shows how for a significant period of PBH domination, a large amount of entropy is dumped into plasma and the thermally produced asymmetry gets diluted. One can analytically quantify this dilution as well.
  Assuming the washout by the inverse decays and PBHs do not coincide in time, the magnitude of the final asymmetry produced by the thermal plasma in presence of PBHs is  then given by
\bea
N_{B-L}^{\rm Final}\simeq N_{B-L}^{\rm Thermal}{\rm Exp}\left[-\int_\infty^0 dz \frac{3}{\tilde{H}z} \frac{\dot{M}_{BH}}{M_{BH}}\frac{\rho_{BH}(z)}{4\rho_{R}(z)}\right].
\eea
 Thus overall, with the  discussion above, we now understand why a long period of PBH domination is preferred to study  a purely non-thermal leptogenesis and as we will  see in the next section, in PBH domination the non-thermally produced  asymmetry is independent of $\beta$.
Let's now calculate the number of a particular species of particles (say $`X$', in our case these are RH neutrinos) emitted by a black hole which is necessary to compute the final baryon asymmetry. The differential number of particles emitted by a black hole can be obtained as\cite{br0a}
\bea
dN=dE/3T_{BH}=\frac{M_{Pl}^2}{24 \pi} \frac{1}{T_{BH}^3}dT_{BH},
\eea
where we have used 
\bea
dE\equiv -d(M_{BH})=\frac{M_{Pl}^2}{8\pi}\frac{dT_{BH}}{T_{BH}^2}
\eea
and consider mean energy of the radiated particles as $\bar{E}=3T$. Assuming the black holes completely evaporate, the total number of $X$ particle emitted can be calculated as
\bea
N_X=\frac{g_X}{g_{*B}}\int_{T_{BH}}^\infty dN=\frac{4\pi}{3}\frac{g_X}{g_{*B}}\left(\frac{M_{BH}}{M_{Pl}}\right)^2 ~~{\rm for}~~T_{BH}>M_X,\\
N_X=\frac{g_X}{g_{*B}}\int_{M_X}^\infty dN=\frac{1}{48\pi}\frac{g_X}{g_{*B}}\left(\frac{M_{Pl}}{M_X}\right)^2~~{\rm for}~~T_{BH}<M_X.
\eea
Having set up all the necessary pre-requisites, we now proceed to the detailed calculation of leptogenesis from PBH evaporation.
\section{Leptogenesis from black hole evaporation in radiation and in black hole domination}\label{sec3}
In this section, we first discuss non-thermal leptogenesis in radiation domination. However, as discussed in the previous section, while in radiation domination, one has to be bit more careful, since the thermal contribution to the lepton asymmetry pops up. Therefore, either one has to distinguish the parameter space of thermal and non-thermal leptogenesis or considering lepton asymmetry produced by PBHs as an additional contribution, re-analyse scenarios of thermal leptogenesis.  That could be interesting to some extent, given the fact that in that case, the predictions of thermal leptogenesis may change significantly. 
\begin{figure}
\includegraphics[scale=.5]{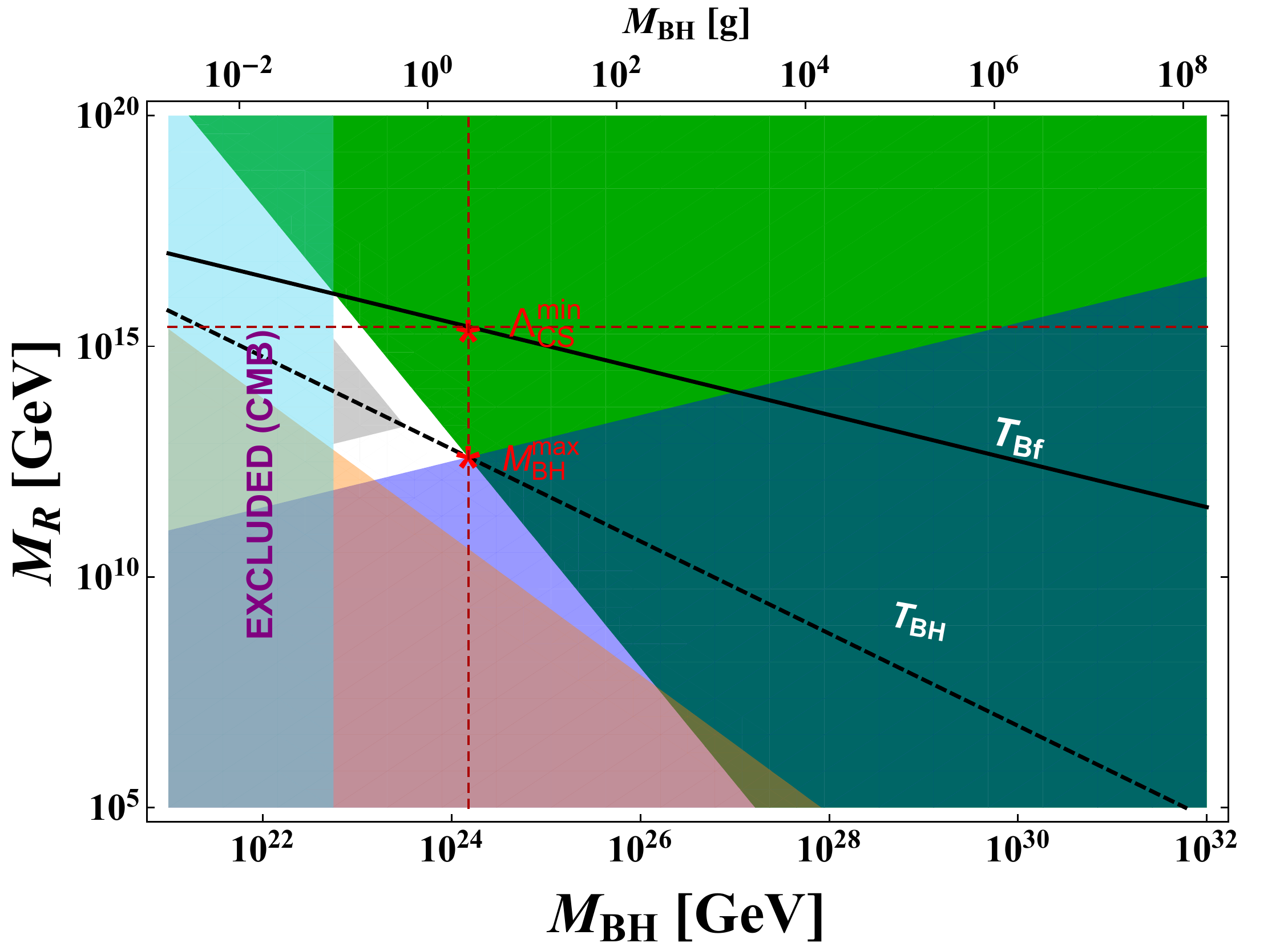}\\
\includegraphics[scale=.5]{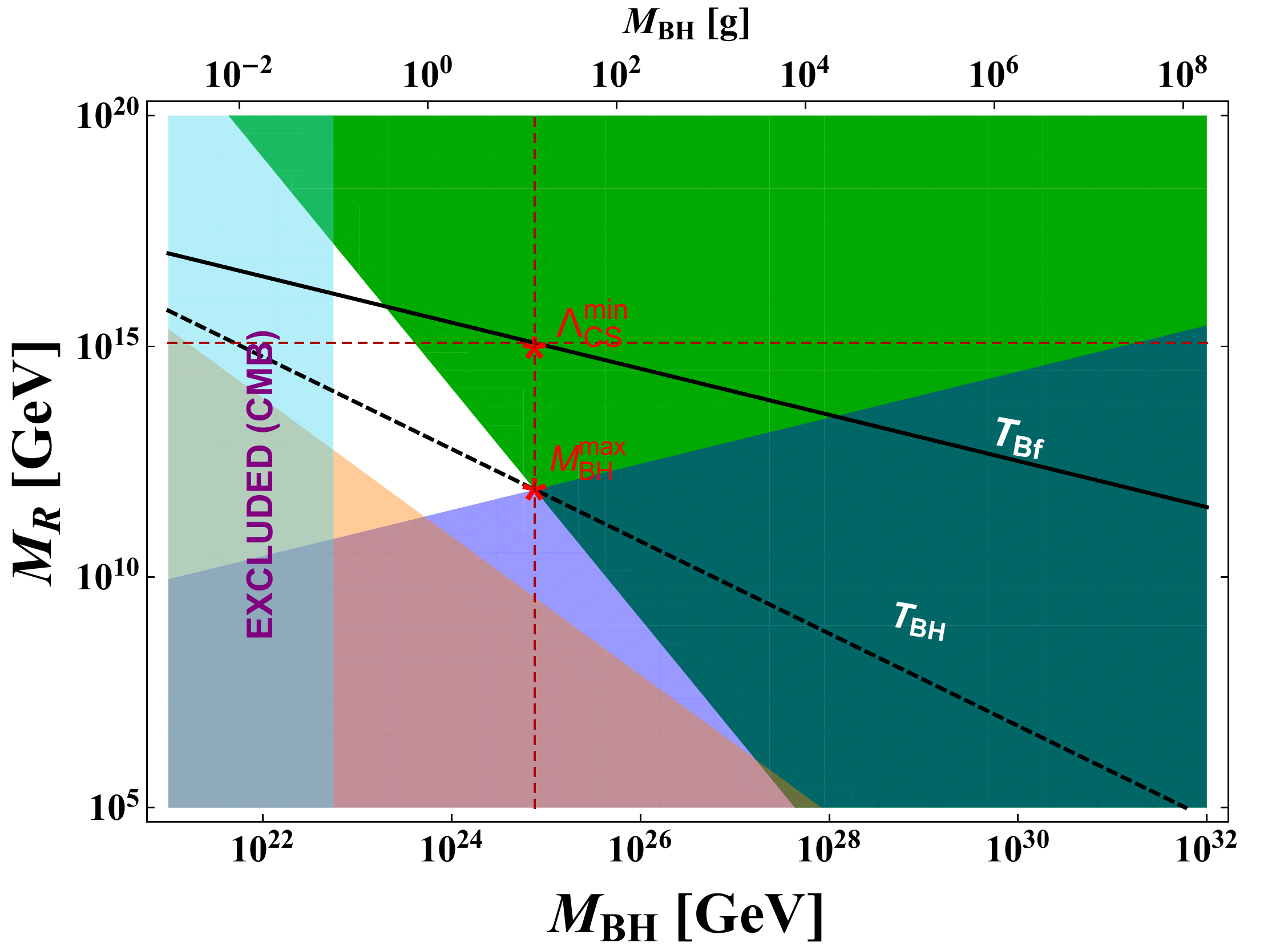}
\caption{Upper Panel: Allowed parameter space for successful leptogenesis in radiation domination (white region) for $\beta=\beta_{\max}\times 10^{-2}$. The same (gray region) for $\beta=\beta_{\max}\times 10^{-1}$. All other regions are excluded by the constraints discussed in the text. Lower panel:  Allowed parameter space for successful leptogenesis in PBH domination (white region).}\label{para}
\end{figure}
Assuming no further entropy production after PBH evaporation, the final asymmetry produced non-thermally by PBH evaporation can be computed as 
\bea
Y_B=a_{\rm sph}N_\nu\varepsilon \frac{n_{BH}(T_{ev})}{s},
\eea
where $a_{\rm sph}\simeq1/3$ is the sphaleron conversion coefficient. Now given the upper bound on the CP asymmetry parameter in Eq.\ref{upep}, one gets the following constraints on the parameter space
\bea
M_R<\left(\frac{m_3M_{Pl}^2}{384 Y_{B}^{\rm Obs}\pi^2v^2}\right)\left(\frac{g_R}{g_{*B}}\right)\frac{n_{BH}(T_{ev})}{s}~~{\rm for}~~M_R>T_{BH},\label{const1}
\eea
and
\bea
1/M_R<\left(\frac{m_3M_{BH}^2}{6Y_{B}^{\rm Obs}M_{Pl}^2v^2}\right)\left(\frac{g_R}{g_{*B}}\right)\frac{n_{BH}(T_{ev})}{s}~~{\rm for}~~M_R<T_{BH},\label{const2}
\eea
where $g_R$ is the number of degrees of freedom associated with the RH neutrino and $Y_B^{\rm Obs}\simeq 8.75\times 10^{-11}$\cite{Planck}. Potentially, there could be another constraint if the RH scale is below the PBH evaporation temperature. In that case, the produced lepton asymmetry is washed out by the inverse decays. This constraint can be worked out after properly computing the inverse decay out-of-equilibrium temperature as\cite{lep4,rome}
\bea
M_R/T^{\rm Inv}_{\rm out}\equiv M_R/T_{ev}>\Phi(K)~~{\rm with}~~\Phi(K_1)=2+4K_1^{0.13}e^{\frac{-2.5}{K_1}}
\eea
which translates into the constraint on $M_R$ as 
\bea
M_R>\left(\frac{45 M_{Pl}^2}{16\pi^3 g_{*}(T_{ev})\tau^2}\right)^{1/4} \Phi(K_1).\label{const3}
\eea
With no fine-tuning in the seesaw formula, $K_1\simeq 10^3m_1$, with $m_1$ being bounded from above from cosmology as $m_1<51$ meV\cite{Planck}. We use this value in the numerical computation.
The  factor  $n_{BH}/s$ factor in radiation domination can be calculated as 
\bea
Y_{BH}^{R}(T_{ev})\equiv \frac{n_{BH}}{s}\left(T_{ev}\right)=\frac{1}{M_{BH}}\frac{\rho_R(T_{ev})}{s(T_{ev})}r(T_{ev}).
\eea
Now using  Eq.\ref{radeng}, Eq.\ref{key}, Eq.\ref{tbf} and Eq.\ref{entr}, $Y_{BH}^{\rm R}(T_{ev})$ can be calculated as
\bea
Y_{ BH}^{\rm R}(T_{ev})=\frac{3\beta}{4}\left(\frac{g_*(T_{Bf})}{g_*(T_{ev})}\right)^{1/4}\frac{g_*(T_{ev})}{g_{*s}(T_{ev})}\left(\frac{45\gamma^2}{16\pi^3 g_*(T_{Bf})}\right)^{1/4}\left(\frac{M_{Pl}}{M_{ BH}}\right)^{3/2}.
\eea
On the other hand, for a significant period of PBH domination, the number density $n_{BH}$ can be calculated as 
\bea
n_{BH}(T_{ev})=\frac{M_{Pl}^2}{6\pi M_{BH}\tau^2},
\eea
where we have assumed $H (T_{ev})\simeq 2/3\tau$.  Now deriving $T_{ev}$ for PBH domination (cf. Eq.\ref{teva} for radiation domination) and using Eq.\ref{lt} as well as Eq.\ref{entr}, one arrives at
\bea
  Y_{ BH}^{\rm B}(T_{ev})=\frac{15}{4\pi^3 g_{*s}(T_{ev})}\left(\frac{\pi^3 g_*(T_{ev})}{5} \right)^{3/4}\left(\frac{\mathcal{G}g_{*B}(T_{ BH})}{10240\pi}\right)^{1/2} \left(\frac{M_{Pl}}{M_{ BH}}\right)^{5/2}.
\eea
In figure \ref{para}, we show the allowed parameter space for successful leptogenesis for radiation (upper panel) and black hole domination (lower panel). In the plots, the green, blue and the red regions are excluded due to the constraints derived in Eq.\ref{const1}, Eq.\ref{const2} and Eq.\ref{const3}.  For radiation domination, the white region is allowed when  we consider $\beta=10^{-1}\beta_{\rm max}$ (cf. Eq.\ref{beta}) while the gray region is allowed for $\beta=10^{-2}\beta_{\rm max}$. We note that the allowed parameter space vanishes for $\beta=10^{-3}\beta_{\rm max}$.  As mentioned earlier, since we are mostly interested in PBH domination, we discuss the parameter space in a bit more detail. The solid black line and the dashed line represent the PBH formation temperature and the Hawking temperature. Therefore, the $U(1)_{B-L}$ breaking scale always lies above the  solid black line. On the other hand, comparing Eq.\ref{const1} and Eq.\ref{const2}, one obtains the upper bound on $M_{BH}$ as
\bea
M_{BH}^{\rm max}=\left[2304\left(\frac{Y_B^{\rm Obs}\pi v^2}{m_3}\right)^2\left(\frac{g_{*B}}{g_R}\right)^2\frac{1}{(\bar{Y}^B_{BH}(T_{ev}))^2}\right]^{-\frac{1}{3}}\simeq 13{\rm g}~(m_3=0.05 {\rm eV}),\label{bhmax}
\eea
where $\bar{Y}^B_{BH}(T_{ev})=Y^B_{BH}(T_{ev})M_{BH}^{5/2}$. Consequently, $\Lambda_{CS}^{\rm min}$ is calculated as 
\bea
\Lambda_{CS}^{\rm min}=\left(\frac{45 \gamma^2}{16\pi^3 g_*(T_{Bf})}\right)^{1/4}\left(\frac{M_{Pl}}{M_{BH}^{\rm max}}\right)^{1/2}M_{Pl}\simeq  1.18\times 10^{15} {\rm GeV}\label{lamcs}
\eea
which can further be  lowered by considering the full range of neutrino $3\sigma$ oscillation data. For smaller $M_{BH}$, one gets $\Lambda_{CS}^{\rm min}$ until the maximum reheat temperature $T_{RH}\sim 10^{16}$ GeV is reached. We take the value of $\Lambda_{CS}^{\rm min}$ in Eq.\ref{lamcs} and compute the GW spectra in the next section. We conclude this section with the following remarks: In the computation of GW spectra from cosmic strings, we do not consider  possible formation of black hole-string networks which for PBHs of solar mass level with $\Lambda_{CS}\gtrsim T_{Bf}$, may leave  observable consequences on the GW spectra\cite{csbl}. Though we are primarily dealing with the non-thermal production, one would like also to understand the overall status of thermal leptogenesis in presence of PBHs. Firstly, as we see from the figure \ref{para}, that the non-thermal contribution becomes prominent for very light black holes and with $M_1\gtrsim 10^{11}$ GeV, whereas for the RH masses, say $M\in [10^9,10^{10}]$, the non-thermal contribution is sub-dominant. However, for $M\in [10^9,10^{10}]$, correct baryon asymmetry can be generated by the standard thermal leptogenesis, since in that case, the light PBHs already evaporate before the asymmetry production from the thermal bath stops. On the left side of figure \ref{fig4}, we show the dynamics of the asymmetry production (standard hierarchical $N_1$-dominated scenario) for $M_{BH}=1$g with $M_1=4\times 10^{12}$ GeV and $M_1=4\times 10^{9}$ GeV. It is clear that for the latter case, the effect of PBH on the final asymmetry is negligible while for the former case, the thermally produced asymmetry is washed out. On the other hand,  heavy PBHs evaporate much later in time and therefore they erase the thermally produced asymmetry unless the typical RH mass scale is below the evaporation temperature. On the right-hand side, we show the dynamics for $M_{BH}=100$g with $M_1=4\times 10^{11}$ GeV and $M_1=4\times 10^{8}$ GeV. This shows that for heavier PBHs, one has to go beyond the hierarchical scenario for a successful thermal leptogenesis, given the fact that for hierarchical leptogenesis, $M_1\gtrsim 10^{9}$ GeV (without flavour effect)\cite{Davidson:2002qv} and $M_1\gtrsim10^{8}$ GeV (with flavour effect)\cite{fle3}. A more precise numerical relation between $M_{BH}$ and the RH scale can be found in Ref.\cite{br4} (which we found in the middle stage of this work) which solves density matrix equations for the lepton asymmetry matrix (treating flavour effects rigorously).
\section{Gravitational waves from cosmic string and  phenomenological discussion}\label{sec4}

\begin{figure}
\includegraphics[scale=.45]{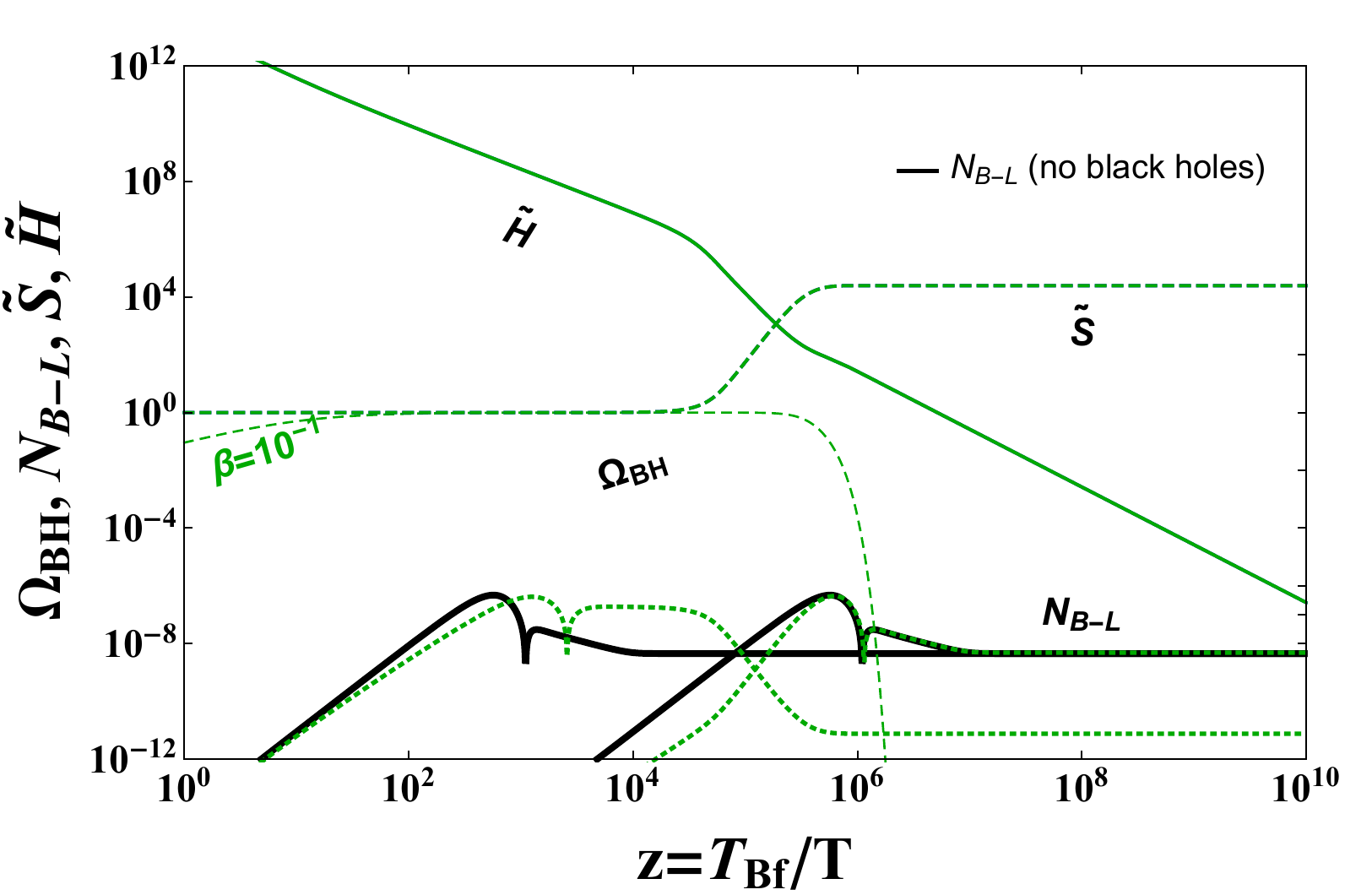} \includegraphics[scale=.45]{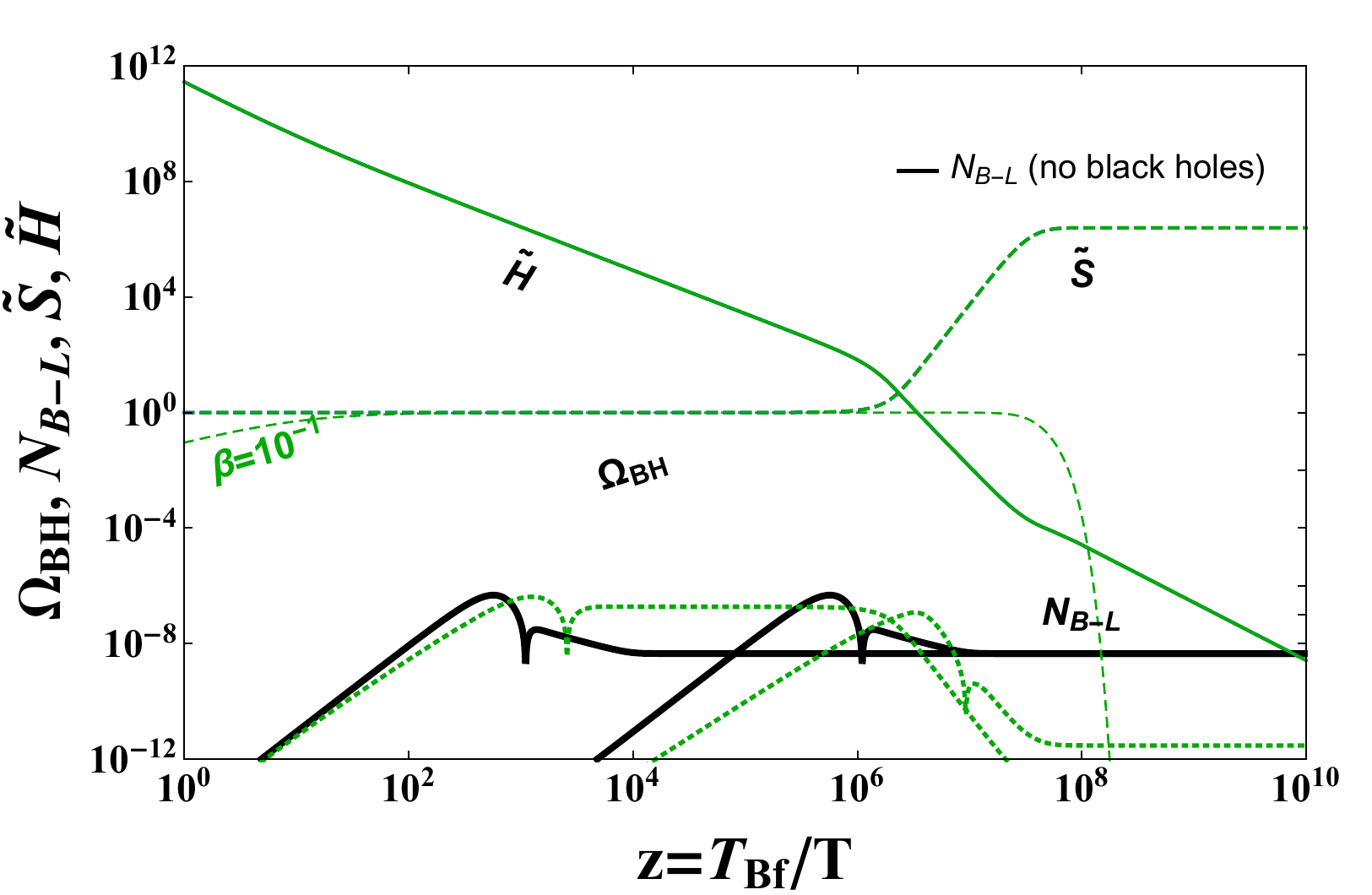}
\caption{Left: Evolution of the $B-L$ asymmetry for $M_{BH}=1$g, $\Delta=10^{-3} (M_1\simeq 4\times 10^{12} {\rm GeV})$ and $\Delta=10^{-6} (M_1\simeq 4\times 10^{9} {\rm GeV})$. Right: Evolution of the $B-L$ asymmetry for $M_{BH}=100$g, $\Delta=10^{-3} (M_1\simeq 4\times 10^{11}{\rm  GeV})$ and $\Delta=10^{-6} (M_1\simeq 4\times 10^{8} {\rm GeV})$.}\label{fig4}
\end{figure}
\begin{figure}
\includegraphics[scale=.44]{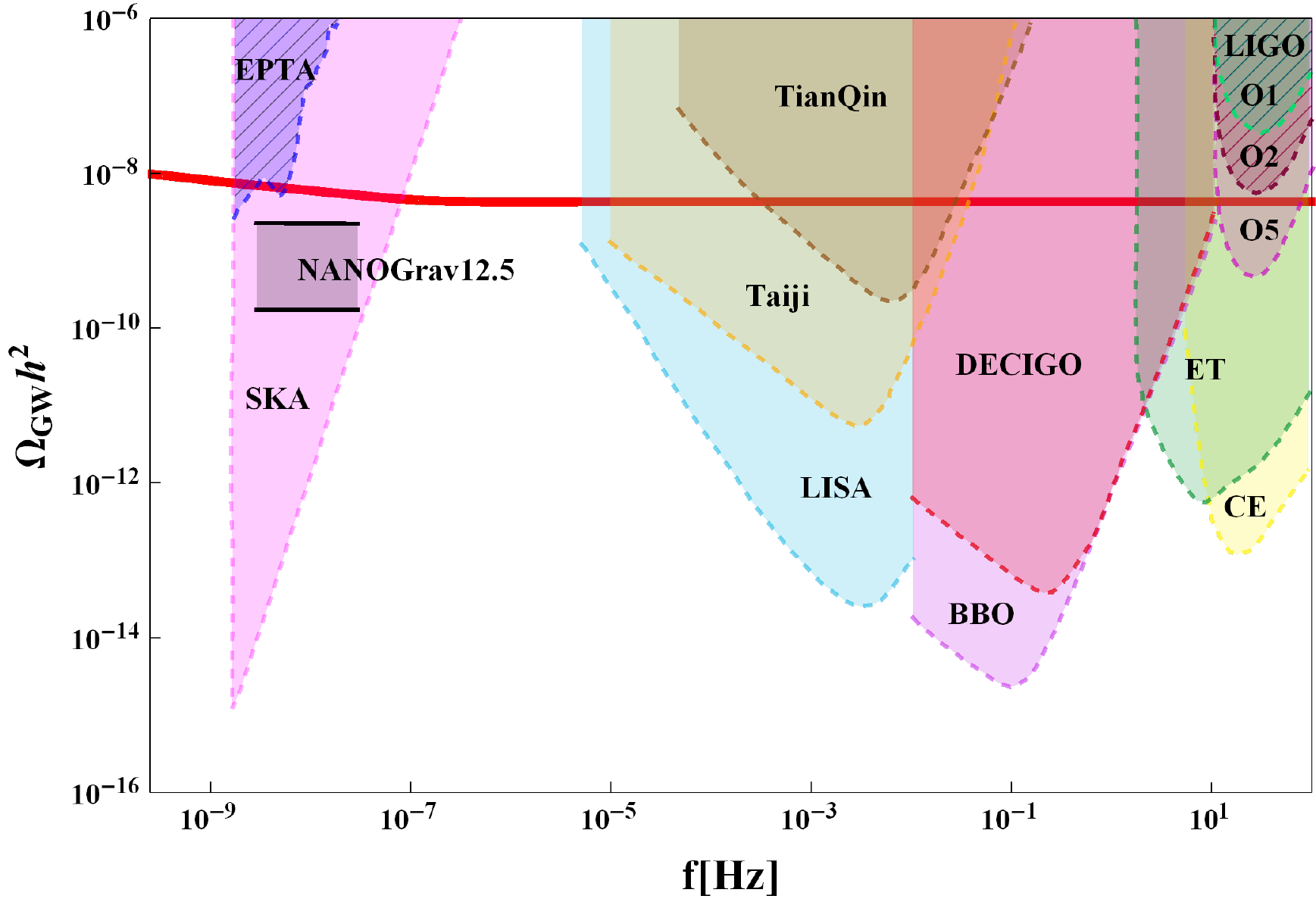}\includegraphics[scale=.44]{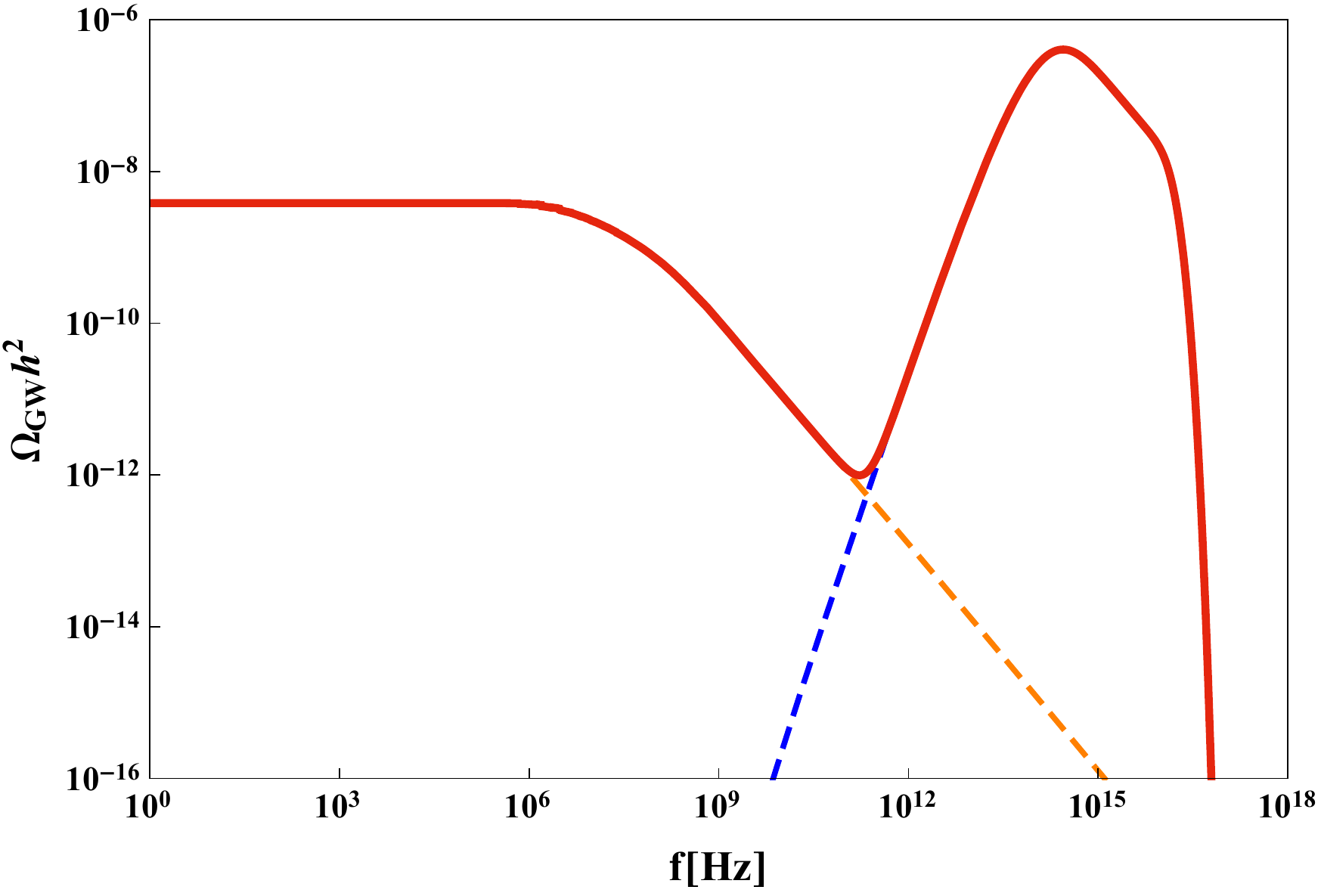}\\
\includegraphics[scale=.44]{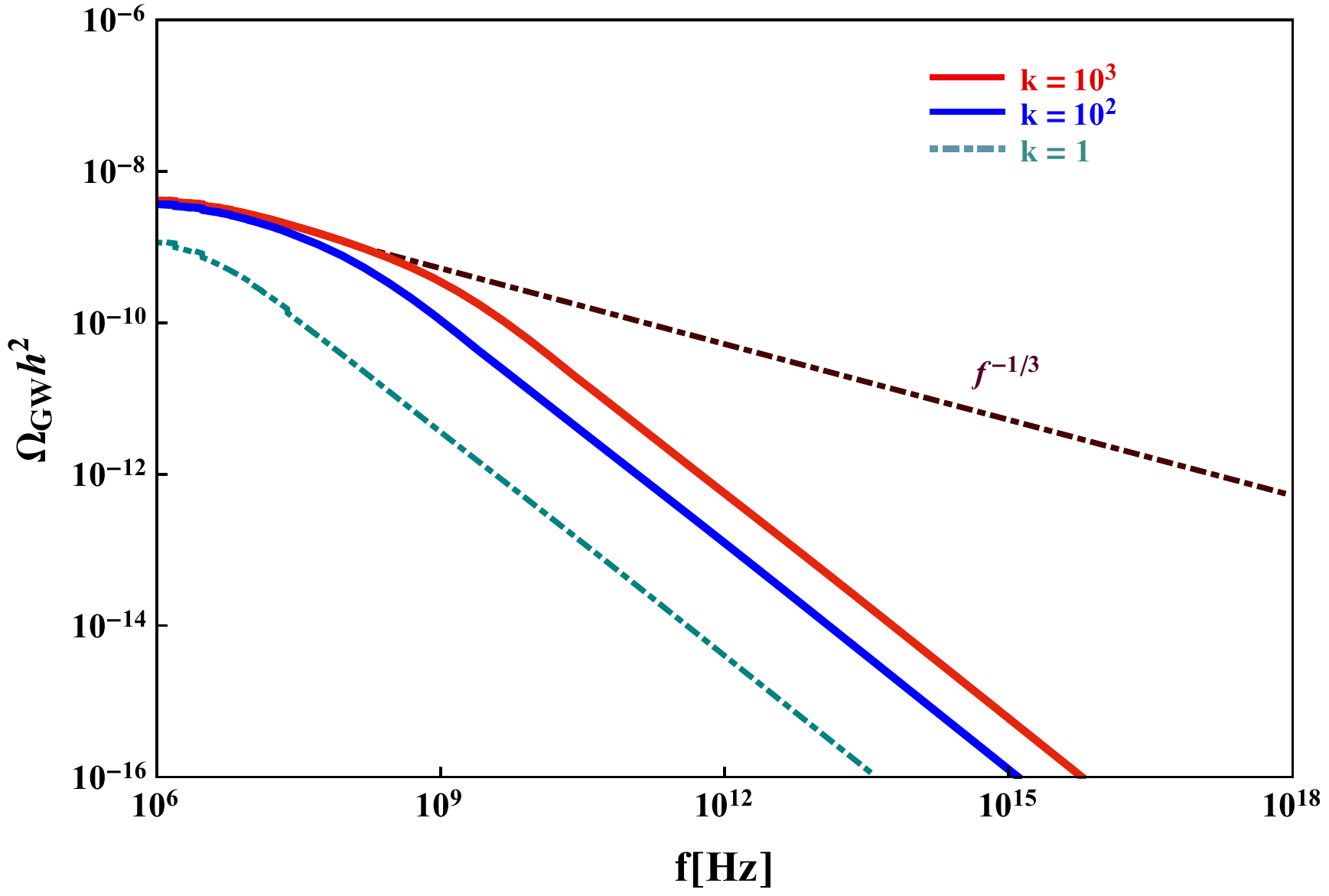}\includegraphics[scale=.44]{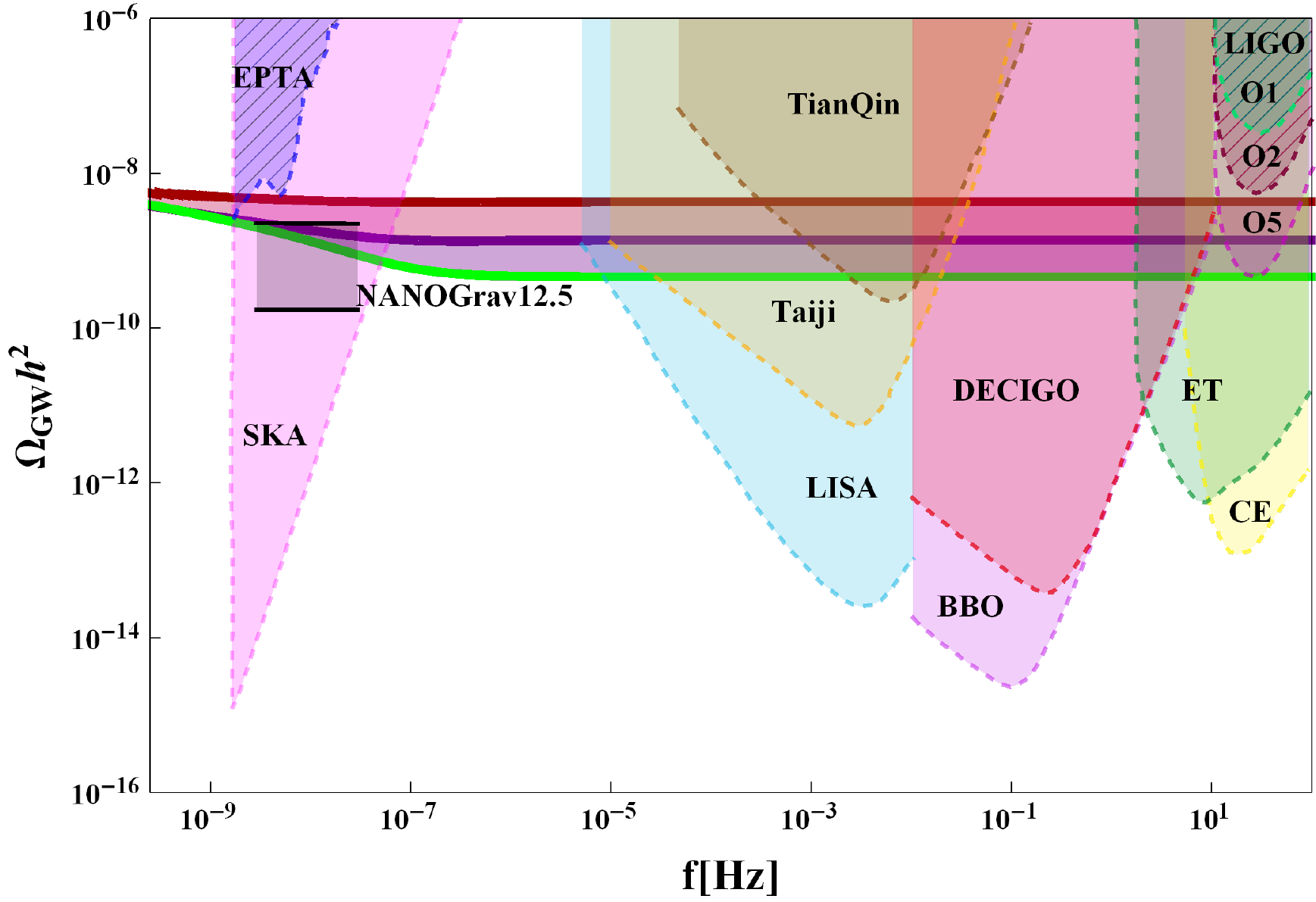}
\caption{Top left: GWs from cosmic strings for $M_{BH}^{\rm max}$ and $\Lambda_{CS}^{\rm min}$ for $\alpha=0.1$. Top right:  GWs from cosmic strings for $\Lambda_{CS}^{\rm min}$ and $\alpha=0.1$ (red dashed) and graviton emission for  $M_{BH}^{\rm max}$ (blue dashed). Bottom left: Effect of including larger number modes on the GW spectrum beyond the turning point frequency. Bottom right: GW spectra for smaller initial loop sizes ($\alpha=10^{-3}, 10^{-4}$ and $ 10^{-5}$) considering the analytic expression obtained from the VOS model.}\label{fig5}
\end{figure}

Now we proceed to the computation of GW spectrum originating from cosmic string network. Cosmic strings  which are nearly one-dimensional objects  arise  generically with  topologically-stable field configurations in BSM theories  featuring spontaneously broken U(1) symmetry 
\cite{cs2,Nielsen:1973cs}. They are considered to be one-dimensional objects characterised macroscopically with their string tension $\mu$ which is typically taken to be of the order of the square of the symmetry breaking scale $\Lambda_{CS}$. Normalised string tension $G\mu$, with $G$ being the Newton's constant, is directly constrained by CMB observation as $G\mu\lesssim  1.1\times 10^{-7}$\cite{stcmb}. After formation, cosmic strings are expected to  reach a scaling regime in which their net energy density tracks the total  energy density of the universe with a relative fraction $G\mu$.  This regime consists of many closed loops as well as Hubble-length long strings which intersect to form new loops as the universe expands. All these loops oscillate and emit radiation, including GWs.  We consider stochastic gravitational background  (SGWB) from cosmic string network by considering Nambu-Goto strings which radiate energy dominantly in the form of GW radiation\cite{ng1,ng2}. We follow Ref.\cite{gr2} to calculate SGWB from cosmic string. Once the loops  are formed, they radiate energy in the form of gravitational radiation at a constant rate as
\bea
\frac{dE}{dt}=-\Gamma G\mu^2,
\eea
where  $\Gamma=50$\cite{Vachaspati:1984gt,Vilenkin:1981bx}.
 Therefore, the initial length $l_i=\alpha t_i$ of the loop decreases as 
\bea
l(t)=\alpha t_i-\Gamma G\mu(t-t_i)
\eea
until it disappears completely. The loop size  $\alpha$ has a distribution and for the largest loop one typically has $\alpha=0.1$\cite{Blanco-Pillado:2017oxo,Blanco-Pillado:2013qja}. We consider $\alpha=0.1$ in the numerical calculation and also discuss the effects of smaller loops in the GW spectrum. The total energy loss from a loop is decomposed
into a set of normal-mode oscillations at frequencies $\tilde{f}=2k/l$, where $k=1,2,3..$. The  GW density parameter is given by
\bea
\Omega_{GW}=\frac{f}{\rho_c}\frac{d\rho_{GW}}{df},
\eea
where $f$ is the red-shifted frequency and $\rho_c=3H_0^2/8\pi G$.  The density parameter $\Omega_{GW}$ can be written as a sum over all relic densities corresponding to a mode $k$ as
\bea
\Omega_{GW}(f)=\sum_k \Omega_{GW}^{(k)}(f),
\eea
where \footnote{Please see Appendix:\ref{ap1} for a derivation using the loop number density obtained  from the Velocity-dependent One-Scale (VOS) model\cite{vos1,vos2,gr3}.}
\bea
\Omega_{GW}^{(k)}(f)=\frac{1}{\rho_c}\frac{2k}{f}\frac{\mathcal{F}_\alpha \Gamma^{(k)} G\mu^2}{\alpha(\alpha+\Gamma G\mu)}\int_{t_F}^{t_0}d\tilde{t}\frac{C_{\rm eff}(t_i^{(k)})}{t_i^{(k)4}}\left[\frac{a(\tilde{t})}{a(t_0)}\right]^5\left[\frac{a(t_i^{k})}{a(\tilde{t})}\right]^3\Theta(t_i^{(k)}-t_F)\label{cseq}
\eea
and the integration runs over the emission time with $t_F$ as the time when the network reach scaling regime shortly  after  formation. The numerical values of $C_{\rm eff}$ are found to be 5.4 and 0.5 at radiation and matter domination and $\mathcal{F}_\alpha$ has a value $\sim 0.1$\cite{Blanco-Pillado:2017oxo,Blanco-Pillado:2013qja}. The quantity $t_i^{(k)}$ is the formation time of the loops contributing to the mode $k$ and is given by
\bea
t_{i}^{(k)}(\tilde{t},f)=\frac{1}{\alpha+\Gamma G \mu}\left[\frac{2k}{f}\frac{a(\tilde{t})}{a(t_0)}+\Gamma G \mu \tilde{t}\right].
\eea
Assuming GW emission is dominated by the cusps like structures, the relative emission rate per mode is given by
\bea
\Gamma^{(k)}=\frac{\Gamma k^{-4/3}}{\sum_{m=1}^\infty m^{-4/3}}
\eea
with $\zeta(4/3)=\sum_{m=1}^\infty m^{-4/3}\simeq 3.6$ and $\sum_k\Gamma^{(k)}=\Gamma$. On the  left side of figure \ref{fig5}, we show the GW spectrum for $\Lambda_{CS}^{\rm min}$ with loop size $\alpha=0.1$ and $k_{max}\simeq 10^3$. As one sees, that even for the $\Lambda_{CS}^{\rm min}$, the mechanism predicts GW with a strong amplitude with a flat plateau  which will be fully tested in  the space-based  interferometers such as  Taiji \cite{taiji}, TianQin\cite{tianquin}, LISA\cite{lisa}, BBO\cite{bbo}, DECIGO\cite{decigo}, ground-based interferometers like Einstein Telescope (ET)\cite{et} and Cosmic Explorer (CE)\cite{ce}, and atomic interferometers MAGIS\cite{magis}, AEDGE\cite{aedge}. 
Note that the amplitude-frequency relation in Eq.\ref{cseq} is explicitly dependent on the cosmological history of the universe. Therefore, any non-standard cosmological scenario (we have a PBH dominated early universe) would leave its imprints on the GW spectra. Any observable break in the GW spectra  due to this non-standard expansion history could be a strong probe of PBH baryogenesis, given that the duration of the non-standard expansion is fixed by the requirement to obtain correct baryon asymmetry. Given $T_\Delta$ is the temperature at which the most recent radiation era begins (before BBN one should have a radiation domination), the frequency (the turning point frequency, see Appendix:\ref{ap2}) at which the flat plateau (see top panel of figure \ref{fig5}) breaks can be estimated as\cite{gr2}
\bea
f_\Delta=\sqrt{\frac{8}{z_{\rm eq}\alpha\Gamma G\mu}}\left[\frac{g_*(T_\Delta)}{g_*(T_0)} \right]^{1/4}\left(\frac{T_\Delta}{T_0}\right)t_0^{-1},\label{break}
\eea
where the red-shift at the standard matter radiation equality is given by $z_{\rm eq}\simeq 3387$ and $T_0=2.7 $K is the present temperature.  For an early matter era, the typical amplitude of the GW for the fundamental mode ($k=1$) goes with frequency as 
\bea
\Omega_{\rm GW}(f)\simeq \Omega_{\rm GW}^{\rm plt}\left(\frac{f}{f_\Delta}\right)^{\kappa},
\eea
where $\kappa\simeq 0$ for $f\lesssim  f_\Delta$ and $\kappa\simeq-1$ for $f\gtrsim f_\Delta $. In the case of PBH baryogenesis $T_{\Delta}\simeq T_{ev}$. Now using Eq.\ref{teva}, Eq.\ref{bhmax} and Eq.\ref{break}, the frequency $f_\Delta$ can be calculated as $f_\Delta\simeq 3\times 10^6$ Hz, for $\alpha=0.1$. Therefore,  to observe this break one needs GW detectors at MHz frequencies. In addition, let us mention that beyond the turning point frequency, the fall of the spectrum as $\Omega_{GW}\propto f^{-1}$ is true only for the fundamental or first few modes\cite{Gouttenoire:2019kij,Cui:2019kkd,lepcs2}. Once larger number of modes are included, the spectrum falls as $f^{-1/3}$ (see Appendix:\ref{ap3} for a derivation with a cartoon diagram)--a feature that can be observed in the bottom-left panel of Fig.\ref{fig5}. There could be another gravitational wave background via graviton emission by the PBHs\cite{Anantua:2008am,Dolgov:2011cq,Dong:2015yjs}. However, for light PBHs as in the present scenario, GW background via graviton emission typically peaks  at frequencies more than $10^{12}$ Hz. An analytical relation between energy density and frequency is given by\cite{Anantua:2008am} 
\bea
\Omega_{GW}^{GR}(f)=\frac{128\pi^4 f^4}{\rho_c M_{Pl}^4}\int_{t_{Bf}}^{t_{ev}}\frac{M(t)^2 n_{PBH}(t)}{\exp\left[16\pi^2 f M(t)a(t_0)/a(t)M_{Pl}^2\right]-1}dt.
\eea
In figure \ref{fig5} (top), we have shown the full spectrum of gravitational waves relevant to probe PBH baryogenesis. In the left panel we have shown the GW spectrum from CS only, for $\alpha=0.1$, whereas in the right panel, we show the GW spectrum from CS for $\alpha=0.1$ (red dashed) as well as due to the graviton emission (blue dashed) for $M_{BH}=13$g. The thick solid red line on the right hand side represents the combined signal that would  manifest a successful baryogenesis.
   There have been proposals to  detect GWs at higher frequencies\cite{Li:2009zzy,Arvanitaki:2012cn,Ito:2019wcb} which may shed light on the behaviour of GW spectrum at higher frequencies compatible with successful PBH baryogenesis. Since we discuss  GWs with strong amplitude, probably it is worthwhile to have some comments on the  prospect of PBH baryogenesis in the light of recent finding by the NANOGrav collaboration\cite{nanog}. The NANOGrav collaboration with their recently released 12.5 yrs data set, reports  strong evidence for a stochastic common-spectrum process  over independent-red noises across 45 pulsars \cite{nanog}.  However, the detection has not been claimed as GWs, since the time residuals do not show characteristic spatial relation which is described by the  Hellings–Downs (HD) curve\cite{Hellings:1983fr}.  On top of that, other systematics such as pulsar spin noise\cite{spin} and solar system effects\cite{solar} may affect the signal and therefore, the analysis requires proper handling of these two effects\cite{nanog}. Nonetheless, if the signal is interpreted as GWs, it would unequivocally open up  a new direction to probe Early Universe cosmology. Note that, the new NANOGrav 12.5 yr data\cite{nanog} is nearly consistent with previous EPTA data\cite{Lentati:2015qwp}, but  they are in  tension with previous limits from PPTA \cite{Shannon:2015ect} and a previous NANOGrav analysis of their 11 yrs (older) data\cite{Arzoumanian:2018saf}. According to Ref.{\cite{nanog}, this tension would be reduced using improved  prior to the intrinsic pulsar red noise to the older data.
The 12.5 yrs NANOGrav data are expressed in terms of power-law signal with characteristic strain given by 
\bea
h_c(f)=A\left(\frac{f}{f_{yr}}\right)^{(3-\gamma)/2}
\eea
with $f_{yr}=1 yr^{-1}$ and $A $ being the characteristic strain amplitude. The abundance of GWs has the standard form and can be recast as:
\bea
\Omega_{GW}(f)=\Omega_{yr}\left(\frac{f}{f_{yr}}\right)^{5-\gamma}, ~~~~~{\rm with}~~~~~ \Omega_{yr}=\frac{2\pi^2}{3H_0^2}A^2f^2_{yr}.\label{po}
\eea
Given our predicted amplitude for $\alpha=0.1$ within the NANOGrav frequency range, we find a fitting point (see Ref.\cite{nan1,lepcs3} for a procedure)   on the  spectral index ($\gamma
$)-amplitude ($A$) plane  well outside the $2\sigma$ contour which also evident from the Fig.\ref{fig5}. It can be seen on the bottom panel in Fig.\ref{fig5}, that the GW amplitude could  be consistent with the NANOGrav data for smaller loop sizes (see also Fig.\ref{fig6} for a fit). As mentioned earlier, recent numerical simulations of cosmic string networks report a population of large loops that effectively contribute to the GWs, with initial loop size parameter peaked around $\alpha\simeq  0.1$\cite{Blanco-Pillado:2017oxo,Blanco-Pillado:2013qja}.  Around 10$\%$ of the energy released by the long string network goes to the large loops, whereas the remaining $90\%$ goes to the kinetic energy of highly boosted smaller loops. The kinetic energy simply redshifts away and is not transferred to GWs. This fact is taken into account in VOS model by the normalization factor $\mathcal{F}_\alpha=0.1$ which for large loops ($\sim \alpha\simeq  0.1$), provides a good description of the results  obtained from numerical simulation  (see Appendix \ref{ap1}). To compute GW spectrum for smaller loops, we use the  loop number density obtained analytically  from VOS model without including the  normalisation factor $\mathcal{F}_\alpha=0.1$ (see e.g., Ref.\cite{gr3}).
\begin{figure}
\includegraphics[scale=.7]{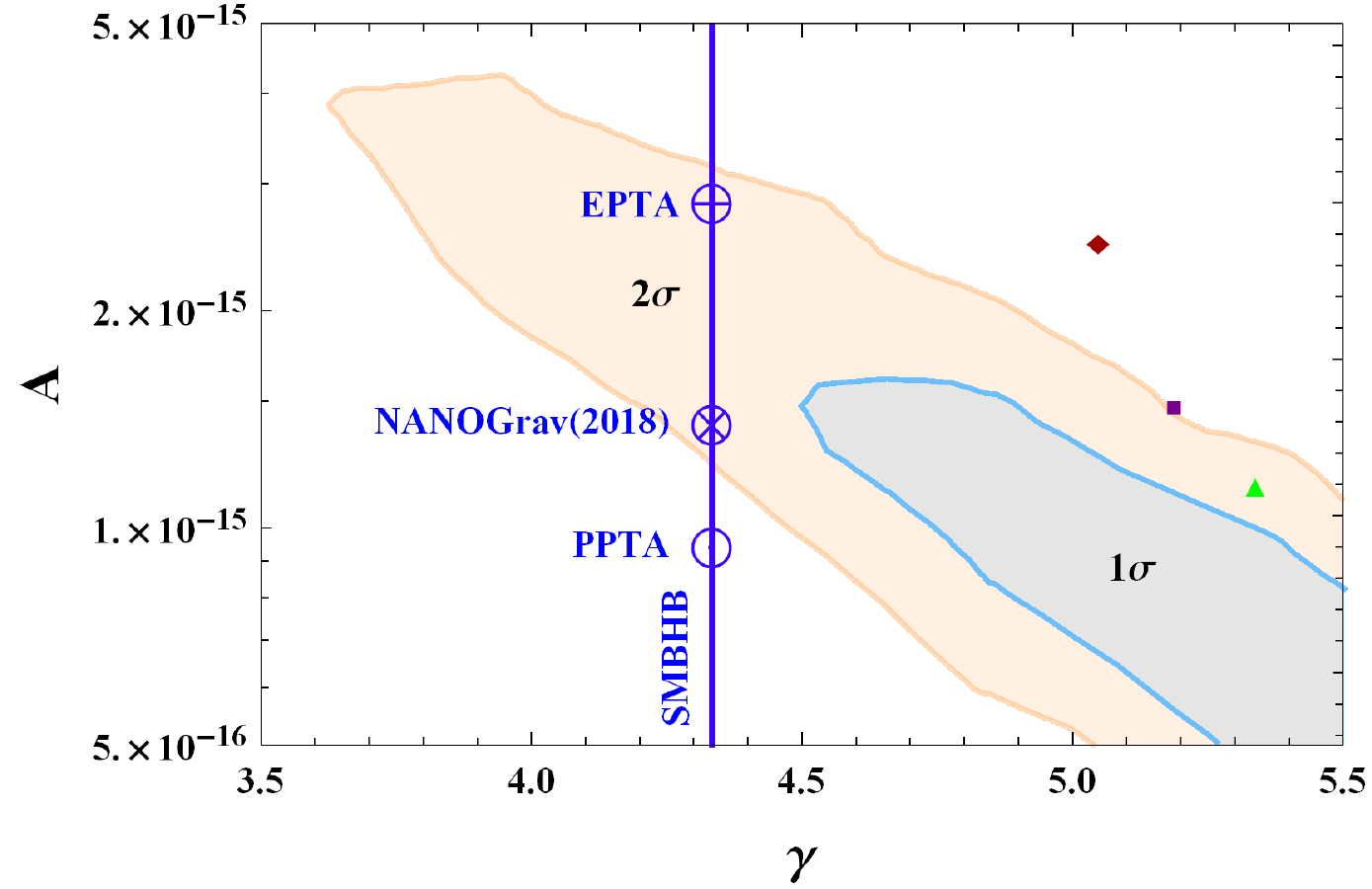}
\caption{A fit to the NANOGrav data for small loops ($\alpha<0.1$ spectrum) and $\mathcal{F}_\alpha=1$. Three benchmark values of $\alpha$ are shown in red ($\alpha=10^{-3}$), purple ($\alpha=10^{-4}$) and green ($\alpha=10^{-5}$).}\label{fig6}
\end{figure}
\section{Summary}\label{sec5}
In this work, we have studied baryogenesis via non-thermal leptogenesis from ultralight primordial black hole evaporation and its test with cosmic string induced gravitational waves. We show that if the right handed neutrino masses are generated dynamically which is then  followed by black hole formation, successful baryogenesis gives rise to GWs with strong amplitude which will be tested within a wide range of frequencies by the future GW detectors. We focus more on the scenario where the black holes dominate the energy density of the universe before they evaporate. In this case, the final baryon asymmetry is independent of black hole density at the formation and there is a break in the GW spectra which occurs around MHz frequency. Therefore, scenarios like PBH baryogenesis could be a good motivation to think of GW detectors beyond the frequency sensitivity of LIGO. We also consider the recent finding  by the NANOGrav PTA of a  stochastic common spectrum process across many pulsars which could be a GW signal. The predicted GW amplitude by successful PBH baryogenesis is in tension  with the  reported cross-power spectral density and characteristic strain amplitude contours for large string loops as predicted by numerical simulations of cosmic string network. However, an agnostic analytical consideration of GWs produced by smaller loops can explain NANOGrav data.
\section*{acknowledgement}
R. Samanta is supported  by the  project  MSCA-IF IV FZU - CZ.02.2.69/0.0/0.0/$20\_079$/0017754 and acknowledges European Structural and Investment Fund and the Czech Ministry of Education, Youth and Sports.
\appendix
\section{Derivation of the Gravitational waves spectrum}\label{ap1}
The normalised energy density of gravitational waves at present time is expressed  as
\bea
\Omega_{GW}(t_0,f)=\frac{f}{\rho_c}\frac{d\rho_{GW}}{df}=\sum_k\Omega_{GW}^{(k)}(t_0,f).
\eea
The frequency derivative is given by\cite{Blanco-Pillado:2013qja}
\bea
\frac{d\rho_{GW}^{(k)}}{df}=\int_{t_F}^{t_0} \left[\frac{a(\tilde{t})}{a(t_0)}\right]^4 P_{GW}(\tilde{t},f_k)\frac{dF}{df}d\tilde{t},\label{a2}
\eea
where the factor $\frac{dF}{df}=f \left[\frac{a(t_0)}{a(\tilde{t})}\right]$ is introduced to take into account the red-shifting of the frequency and the quantity $P_{GW}(\tilde{t},f_k)$ represents the power emitted by the loops and is given by
\bea
P_{GW}(\tilde{t},f_k)=G\mu^2\Gamma_k\int  n(l,\tilde{t}) \delta\left(f_k-\frac{2k}{l}\right) dl.\label{a3}
\eea
Integrating Eq.\ref{a3} over the loop lengths gives 
\bea
P_{GW}(\tilde{t},f_k)=\frac{2kG\mu^2 \Gamma_k}{f_k^2} n(\tilde{t},f_k)=\frac{2kG\mu^2 \Gamma_k}{f^2\left[\frac{a(t_0)}{a(\tilde{t})}\right]^2}n\left(\tilde{t},\frac{2k}{f}\left[\frac{a(\tilde{t})}{a(t_0)}\right]\right)\label{a4}
\eea
From Eq.\ref{a4} and Eq.\ref{a2} one gets
\bea
\frac{d\rho_{GW}^{(k)}}{df}=\frac{2kG\mu^2 \Gamma_k}{f^2}\int_{t_F}^{t_0} \left[\frac{a(\tilde{t})}{a(t_0)}\right]^5 n\left(\tilde{t},\frac{2k}{f}\left[\frac{a(\tilde{t})}{a(t_0)}\right]\right)d\tilde{t}
\eea
Therefore the energy density corresponding to the mode `$k$' is given by
\bea
\Omega_{GW}^{(k)}(t_0,f)=\frac{2kG\mu^2 \Gamma_k}{f\rho_c}\int_{t_F}^{t_0} \left[\frac{a(\tilde{t})}{a(t_0)}\right]^5 n\left(\tilde{t},\frac{2k}{f}\left[\frac{a(\tilde{t})}{a(t_0)}\right]\right)d\tilde{t}.\label{form1}
\eea
The most important aspect to obtain gravitational wave spectrum from cosmic string loops is to compute the number density $n\left(\tilde{t},l_k(\tilde{t})\equiv \frac{2k}{f}\left[\frac{a(\tilde{t})}{a(t_0)}\right]\right)$. In this regard, two different approaches can be considered. The first approach could be  obtaining the number density directly from numerical solution\cite{Blanco-Pillado:2017oxo} which reads (considering the loops created during radiation domination, similar argument also holds true for matter domination)
\bea
n(\tilde{t},l_{k}(\tilde{t}))=\frac{0.18}{\left[l_k(\tilde{t})+\Gamma G \mu\tilde{t}\right]^{5/2}\tilde{t}^{3/2}}.\label{nub}
\eea
On the other hand one may opt for an analytic approach by considering the Velocity dependent One Scale (VOS) model which gives
\bea
n(\tilde{t},l_{k}(\tilde{t}))=\frac{A_r}{\alpha}\frac{(\alpha+\Gamma G \mu)^{3/2}}{\left[l_k(\tilde{t})+\Gamma G \mu\tilde{t}\right]^{5/2}\tilde{t}^{3/2}}\equiv\frac{A_r N_\alpha}{\left[l_k(\tilde{t})+\Gamma G \mu\tilde{t}\right]^{5/2}\tilde{t}^{3/2}},\label{vos}
\eea
where $A_r=5.4$\cite{gr3}. Note that the  VOS model assumes all the loops are of same length at creation. However, at the moment of production, the loops may follow a distribution depending on $\alpha$. In that case the above formula should be modified as 
\bea
n(\tilde{t},l_{k}(\tilde{t}))=\frac{A_r \int w(\alpha) N_\alpha d\alpha}{\left[l_k(\tilde{t})+\Gamma G \mu\tilde{t}\right]^{5/2}\tilde{t}^{3/2}}.
\eea
Therefore, to use the VOS formula in Eq.\ref{vos} and at the same time to be consistent with the numerical result, one has to normalise Eq.\ref{vos}, i.e., 
\bea
n(\tilde{t},l_{k}(\tilde{t}))=\frac{\mathcal{F}_\alpha A_r N_\alpha}{\left[l_k(\tilde{t})+\Gamma G \mu\tilde{t}\right]^{5/2}\tilde{t}^{3/2}},~~{\rm with}~~\mathcal{F}_\alpha =N_\alpha ^{-1}\int w(\alpha) N_\alpha d\alpha
\label{mvos}
\eea
As one can see that for $\alpha=0.1$, Eq.\ref{nub} and Eq.\ref{vos} is consistent for $\mathcal{F}_\alpha\sim 0.18/( A_r \sqrt{\alpha})\sim 0.1$ which we consider in the numerical computation. The most general formula for the number density in an expanding background that scales as $a\sim t^\beta$ is given by
\bea
n(\tilde{t},l_{k}(\tilde{t}))=\frac{A_\beta}{\alpha}\frac{(\alpha+\Gamma G \mu)^{3(1-\beta)}}{\left[l_k(\tilde{t})+\Gamma G \mu\tilde{t}\right]^{4-3\beta}\tilde{t}^{3\beta}}.\label{genn}
\eea
Now having initial time $t_i^{(k)}$ as
\bea
t_i^{(k)}=\frac{l_k(\tilde{t})+\Gamma G\mu \tilde{t}}{\alpha+\Gamma G\mu},
\eea
Eq.\ref{genn} can be re-expressed as
\bea
n(\tilde{t},l_{k}(\tilde{t}))=\frac{A_\beta(t_i^{(k)})}{\alpha(\alpha+\Gamma G \mu)t_i^{(k)4}}\left[\frac{a(t_i^{k})}{a(\tilde{t})}\right]^3.\label{init}
\eea
Putting the value of $n(\tilde{t},l_{k}(\tilde{t}))$ from Eq.\ref{init} into Eq.\ref{form1}, one gets 
\bea
\Omega_{GW}^{(k)}(t_0,f)=\frac{2kG\mu^2 \Gamma_k}{f\rho_c \alpha(\alpha+\Gamma G \mu)}\int_{t_F}^{t_0} \left[\frac{a(\tilde{t})}{a(t_0)}\right]^5\frac{C_{\rm eff}(t_i^{(k)})}{t_i^{(k)4}} \left[\frac{a(t_i^{k})}{a(\tilde{t})}\right]^3 \Theta(t_i^{(k)}-t_F)d\tilde{t},\label{form2}
\eea
where we have renamed $A_\beta$ as $C_{\rm eff}$ and the quantity $\Gamma^{(k)}$ is given by
\bea
\Gamma^{(k)}=\frac{\Gamma k^{-\delta}}{\sum_{m=1}^\infty m^{-\delta}}, ~~\delta=\begin{cases}4/3 ~~{\rm cusps}\\5/3 ~~{\rm kinks}. 
\end{cases}
\eea
\section{Derivation of the flat plateau and the turning point frequency}\label{ap2}
The expression for the fundamental mode in Eq.\ref{form1} can be re-written as 
\bea
\Omega_{GW}^{(k)}(t_0,f)=\frac{2G\mu^2 \Gamma}{f\zeta(\delta)\rho_c}\int_{t_F}^{t_0} \left[\frac{a(\tilde{t})}{a(t_0)}\right]^5 n\left(\tilde{t},\frac{2k}{f}\left[\frac{a(\tilde{t})}{a(t_0)}\right]\right)d\tilde{t}.\label{form3}
\eea
It is convenient to do the integration with respect to the scale factor. Eq.\ref{form3} then becomes
\bea
\Omega_{GW}^{(k)}(t_0,f)=\frac{16\pi}{3\zeta(\delta)}\left(\frac{G\mu}{H_0}\right)^2\frac{\Gamma}{f a(t_0)}\int_{a_F}^{a_{eq}}H(a)^{-1} \left[\frac{a(\tilde{t})}{a(t_0)}\right]^4 n\left(\tilde{t},\frac{2k}{f}\left[\frac{a(\tilde{t})}{a(t_0)}\right]\right)da,\label{form4}
\eea
where
\bea
H=H_0\Omega_r^{1/2}\left(\frac{a(\tilde{t})}{a(t_0)}\right)^{-2}~~{\rm with}~~\Omega_r\simeq 9\times10^{-5}.
\eea
The number density $n\left(\tilde{t},l_k(\tilde{t})\equiv \frac{2k}{f}\left[\frac{a(\tilde{t})}{a(t_0)}\right]\right)$ in Eq.\ref{vos} can also be expressed in terms of the scale factor and is given by
\bea
n(\tilde{t},l_{k}(\tilde{t}))=\frac{A_r}{\alpha}\frac{(\alpha+\Gamma G \mu)^{3/2}}{\left[\frac{2}{f}\left[\frac{a(\tilde{t})}{a(t_0)}\right]+\Gamma G \mu/2H\right]^{5/2}(2H)^{-3/2}.}\label{numrad}
\eea
Putting Eq.\ref{numrad} in Eq.\ref{form4} and after performing the integration one gets 
\bea
\Omega_{GW}^{(1)}(f)=\frac{128\pi G\mu}{9\zeta(\delta)}\frac{A_r}{\epsilon_r}\Omega_r(1+\epsilon_r)^{3/2}\left[\left( \frac{f}{f+\epsilon_r f_{\rm min}\left(\frac{t_i}{t_{\rm eq}}\right)^{1/2}}\right)^{3/2}-\left( \frac{f}{f+\epsilon_r f_{\rm min}}\right)^{3/2}\right],
\eea
where we define $\epsilon_r=\alpha/\Gamma G \mu$ the $f_{min}=\frac{2}{\alpha t_i}\frac{a_i}{a_0}=\frac{4H_0\Omega_r^{1/2}}{\alpha}\frac{a_0}{a_i}$ is the minimum frequency emitted by  a given loop.  Now given the scaling solution of the loop production rate, which decreases with the fourth power in time,  $f\simeq f_{\rm min}$ is  a reasonable assumption. In that case, with $t_i\ll t_{\rm eq}$ one has 
\bea
\Omega_{GW}^{(1)}(f)=\frac{128\pi G\mu}{9\zeta(\delta)}\frac{A_r}{\epsilon_r}\Omega_r\left[(1+\epsilon_r)^{3/2}-1\right]. \label{flp}
\eea
Note that the expression for the flat plateau matches with Ref.\cite{gr3} apart from the factor $\zeta(\delta)$ in the denominator. This is simply due to the fact that our definition of $\Omega_{GW}^{(k)}(f)$ in Eq.\ref{form2} is inclusive of $\Gamma^{k}$. In the above, we have assumed that the dominant emission comes from the very earliest epoch of loop creation.  In practice, a precise value of the time can be computed by maximizing the integral in Eq.\ref{form3} with respect to $\tilde{t}$. This gives 
\bea
\tilde{t}_M\simeq\frac{2}{f\Gamma G\mu}\frac{a_M}{a_0}\equiv \frac{1}{2\Gamma G\mu}\frac{4 a_M}{f a_0}\equiv \frac{l_i}{2\Gamma G\mu}\label{half}
\eea
where $f$ is the frequency observed today that was emitted at time $t_M$ when the a given loop $l_i=\alpha t_i$ reached to the half of its size $l_i/2$, i.e., $t_M$ is eventually the half life of a given loop. Given the time $t_\Delta$ at which the most recent radiation domination begins, the approximate  frequency up to which the spectrum shows a flat plateau is given by
\bea
f_\Delta=\sqrt{\frac{8}{\Gamma G\mu}}t_\Delta^{-1/2}t_0^{-2/3}t_{\rm eq}^{1/6}\simeq \sqrt{\frac{8 z_{\rm eq}}{\Gamma G\mu}}\left(\frac{t_{\rm eq}}{t_\Delta}\right)^{1/2}t_0^{-1}.\label{br0}
\eea
Eq.\ref{br0} when expressed in terms of temperature gives Eq.\ref{break}.
\section{Effect of large-k modes on the spectral shape beyond the turning point}\label{ap3}
\begin{figure}
\includegraphics[scale=.45]{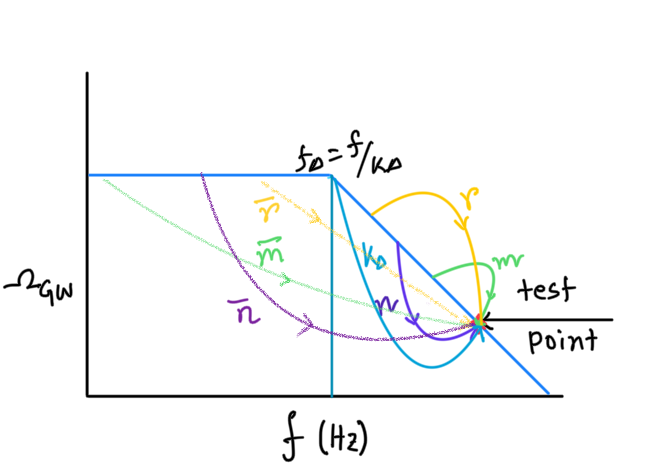}
\caption{A cartoon diagram showing how the contributions from large-k modes  are accounted for to analyse spectral shape beyond the turning point.}\label{figca}
\end{figure}
As pointed out in sec.\ref{sec4}, when the number of modes increases in the sum, the spectral behaviour  beyond the turning point deviates from that of the fundamental mode\cite{Gouttenoire:2019kij,Cui:2019kkd,lepcs2}. The reason being the following: \\

First of all,  from Eq.\ref{form2} it is evident that 
\bea
\Omega_{GW}(f)=\sum_k\Omega_{GW}^{(k)}(f)=\sum_k k^{-\delta}\Omega^{(1)}(f/k).\label{sumk}
\eea
Now to perform the sum or in other words to compute the GW amplitude at a given frequency (called test point in the cartoon diagram (Fig.\ref{figca})) let us expand the RHS of Eq.\ref{sumk} for some first few benchmark modes, i.e.,
\bea
\Omega_{GW}(f)&=&\sum_k k^{-\delta}\Omega^{(1)}(f/k) \nonumber \\ &=& 1^{-\delta}\Omega^{(1)}(f/1)+ m^{-\delta}\Omega^{(1)}(f/m)+ n^{-\delta}\Omega^{(1)}(f/n)+ r^{-\delta}\Omega^{(1)}(f/r)+...,
\eea
where the integers obey $1<m<n<r$. This suggests, if we keep on increasing the mode numbers, there will be critical value  $k\equiv k_\Delta$ for which the amplitude $\Omega_{GW}^{(1)}(f_\Delta=f/k_\Delta)$ contributes to the test point. Therefore we can split the sum into two parts. The first one is from $k=1$ up to $k_\Delta$ for which the test point receives contributions from the non-flat part of the fundamental spectrum and the second one is from $k_\Delta$ to $k_{max} (k_\Delta<\bar{r}<\bar{n}<\bar{m}<k_{max})$ for which the test point receives contribution from the flat part of the fundamental spectrum, i.e., 
\bea
\Omega_{GW}(f)&=&\sum_{k=1}^{k=k_{\Delta}}k^{-\delta}\Omega_{GW}^{(1)}(f/k>f_\Delta)+\sum_{k=k_{\Delta}}^{k=k_{max}}k^{-\delta}\Omega_{GW}^{(1)}(f/k<f_\Delta)\\
&=&\sum_{k=1}^{k=k_{\Delta}}k^{-\delta}\Omega_{GW}^{\rm plt}\left(\frac{f_\Delta}{f/k}\right)+\sum_{k=k_{\Delta}}^{k=k_{max}}k^{-\delta}\Omega_{GW}^{\rm plt}.\label{2sum}
\eea
This sum can easily be evaluated in the large $k_\Delta$ limit using asymptotic expansion of Euler-Maclaurin series for the first term and Hurwitz zeta function for the second term. The final result is  (up to some pre-factor depending on $\delta$)
\bea
\Omega_{GW}(f)\simeq \Omega_{GW}^{\rm plt} \left( \frac{f_\Delta}{f}\right)^{\delta-1}~~{\rm i.e.,}~~\Omega_{GW}(f)\propto\begin{cases}f^{-1/3} ~~{\rm cusps}\\f^{-2/3}~~{\rm kinks}. \end{cases}
\eea
As an aside, let us mention that in the scenarios like kination\cite{w1a,w1b} where the equation of state is stiffer ($w=1$), beyond the turning point, the energy density goes as $f^1$\cite{gr2}. In that case, the first term (swapping  numerator and denominator) in Eq.\ref{2sum} gives the dominant contribution. In the large $k_\Delta$ limit, the sum is therefore
\bea
\Omega_{GW}(f)\simeq  \Omega_{GW}^{\rm plt} \left(\frac{f}{f_\Delta} \right)\sum_{k=1}^{k=k_{\Delta}}k^{-(\delta +1)}\simeq \Omega_{GW}^{\rm plt} \left(\frac{f}{f_\Delta} \right)\zeta(\delta + 1).
\eea
Thus for  stiffer equation of states like kination, the spectral shape is quite similar to the fundamental mode even after taking in account the contribution from the higher k-modes.

{}
  
\end{document}